\documentclass[preprint,10pt,number,5p,twocolumn]{elsarticle}

\usepackage{amssymb}

\usepackage{amsmath}
\usepackage{booktabs}
\usepackage[x11names,table,xcdraw]{xcolor}
\usepackage{longtable}
\usepackage{adjustbox}
\usepackage{makecell}
\usepackage{subfigure}
\usepackage{lipsum}
\usepackage{pifont}
\usepackage{tabularx}
\usepackage{ctable}
\usepackage{pbox}
\usepackage[T1,hyphens]{url}  
\usepackage{multirow}
\usepackage{makecell}
\usepackage{hhline}
\usepackage{wasysym}
\usepackage{amssymb}
\usepackage[acronym,smallcaps]{glossaries}
\usepackage[normalem]{ulem}
\useunder{\uline}{\ul}{}

\usepackage{array, booktabs}
\usepackage{graphicx}

\newcommand{\foo}{\color{LightSteelBlue3}\makebox[0pt]{\textbullet}\hskip-0.5pt\vrule width 1pt\hspace{\labelsep}}

\newacronym{ai}{AI}{Artificial Intelligent}
\newacronym{alto}{ALTO}{Application-Layer Traffic Optimization}
\newacronym{wan}{WAN}{Wide Area Network}
\newacronym{asic}{ASIC}{Application Specific Integrated Circuits}
\newacronym{osi}{OSI}{Open Source Initiative}
\newacronym{bsd}{BSD}{Berkeley Software Distribution}
\newacronym{sriov}{SR-IOV}{Single Root I/O Virtualization}
\newacronym{ict}{ICT}{Information and Communication Technology}

\newacronym{epa}{EPA}{Enhanced Platform Awareness}
\newacronym{numa}{NUMA}{Non-Uniform Memory Access}

\newacronym{enb}{eNB}{Evolved NodeB}
\newacronym{rsu}{RSU}{Roadside Unit}

\newacronym{escape}{ESCAPE}{Extensible Service ChAin Prototyping
Environment}

\newacronym{pox}{POX}{Python-based software-defined networking}

\newacronym{iot}{IoT}{Internet of Things}

\newacronym{gui}{GUI}{Graphical User Interface}
\newacronym{odl}{ODL}{OpenDaylight}
\newacronym{os}{OS}{OpenStack}

\newacronym{dpdk}{DPDK}{Data Plane Development Kit}
\newacronym{xdpd}{xDPd}{eXtensible OpenFlow DataPath daemon}

\newacronym{lsi}{LSI}{Logical Switch Instance}

\newacronym{cots}{COTS}{Commodity Off the Shelf}

\newacronym{capex}{CAPEX}{Capital Expenditure}
\newacronym{opex}{OPEX}{Operational Expenditure}

\newacronym{it}{IT}{Information Technologies}
\newacronym{ha}{HA}{High-Availability}
\newacronym{nsc}{NSC}{Network Service Chain}

\newacronym{em}{EM}{Element Management}
\newacronym{ems}{EMS}{Element Management Systems}
\newacronym{oss}{OSS}{Operation Support Systems}
\newacronym{bss}{BSS}{Business Support Systems}
\newacronym{oam}{OAM}{Operations and Management}

\newacronym[longplural=Points of Presence]{pop}{PoP}{Point of Presence} 

\newacronym{abno}{ABNO}{Application-based Network Operations}
\newacronym{cord}{CORD}{Central Office Re-architected as a Datacenter}

\newacronym{api}{API}{Application Programming Interface} 

\newacronym{cli}{CLI}{Command Line Interface} 

\newacronym{sdo}{SDO}{Standards Developing Organization}

\newacronym{hw}{HW}{hardware} 

\newacronym{ott}{OTT}{Over The Top}

\newacronym{sp}{SP}{Service Provider}

\newacronym{xacml}{XACML}{eXtensible Access Control Markup Language}
\newacronym{pap}{PAP}{Policy Administration Point}
\newacronym{pep}{PEP}{Policy Enforcement Point}
\newacronym{pdp}{PDP}{Policy Decision Point}
\newacronym{pip}{PIP}{Policy Information Point}

\newacronym{nat}{NAT}{Network Address Translation}
\newacronym{dhcp}{DHCP}{Dynamic Host Configuration Protocol}
\newacronym{bng}{BNG}{Broadband Network Gateway}
\newacronym{fw}{FW}{Firewall}
\newacronym{vfw}{vFW}{virtual Firewall}
\newacronym{lb}{LB}{Load Balancer}

\newacronym{slm}{SLM}{Service Level Measurement}

\newacronym{sla}{SLA}{Service Level Agreement}

\newacronym{kqi}{KQI}{Key Quality Indicator}

\newacronym{kpi}{KPI}{Key Performance Indicator}

\newacronym{qos}{QoS}{Quality of Service}

\newacronym{vm}{VM}{Virtual Machine}

\newacronym{dpi}{DPI}{Deep Packet Inspection}
\newacronym{vdpi}{vDPI}{virtual Deep Packet Inspection}

\newacronym{cp}{CP}{Control Plane}
\newacronym{dp}{DP}{Data Plane}

\newacronym{soa}{SOA}{Service-Oriented Architecture}

\newacronym{ietf}{IETF}{Internet Engineering Task Force}
\newacronym{irtf}{IRTF}{Internet Research Task Force}
\newacronym{itu}{ITU}{International Telecommunication Union}
\newacronym{mef}{MEF}{Metro Ethernet Forum}
\newacronym{nist}{NIST}{National Institute of Standards and Technology}
\newacronym{3gpp}{3GPP}{3rd Generation Partnership Project}

\newacronym{fi}{FI}{Future Internet}

\newacronym{do}{DO}{Domain Orchestrator}

\newacronym[longplural=Multi-Domain Orchestrators]{mdo}{MDO}{Multi-Domain Orchestrator}

\newacronym{saas}{S$aa$S}{Software as a Service}
\newacronym{iaas}{I$aa$S}{Infrastructure as a Service}
\newacronym{paas}{P$aa$S}{Platform as a Service}
\newacronym{naas}{N$aa$S}{Network as a Service}
\newacronym{nfviaas}{NFVI$aa$S}{\gls{nfvi} as a Service}
\newacronym{vnfaas}{VNF$aa$S}{\gls{vnf} as a Service}
\newacronym{slaas}{Sl$aa$S}{Slice as a Service}
\newacronym{vnpaas}{VNP$aa$S}{Virtual Network Platform as a Service}
\newacronym{xaas}{X$aa$S}{Anything as a Service}

\newacronym{etsi}{ETSI}{European Telecommunications Standards Institute}
\newacronym{isg}{ISG}{Industry Specification Group}

\newacronym[longplural=Network Function
Virtualization]{nfv}{NFV}{Network Function Virtualization}

\newacronym{vnf}{VNF}{Virtualized Network Function}

\newacronym{pnf}{PNF}{Physical Network Function}

\newacronym{vdu}{VDU}{Virtual Deployment Unit}

\newacronym{vnfd}{VNFD}{Virtualized Network Function Descriptor}

\newacronym{nfvo}{NFVO}{Network Function Virtualization Orchestrator}
\newacronym{vnfm}{VNFM}{VNF Manager}
\newacronym{vim}{VIM}{Virtualized Infrastructure Manager}

\newacronym{nfvi}{NFVI}{NFV Infrastructure}

\newacronym{nf}{NF}{Network Function}

\newacronym{vl}{VL}{Virtual Link}

\newacronym{vnffg}{VNF-FG}{\gls{vnf} Forwarding Graph}

\newacronym{mano}{MANO}{Management and Orchestration}

\newacronym{nfvmano}{NFV-MANO}{NFV-Management and Orchestration}

\newacronym{nso}{NSO}{Network Service Orchestration}

\newacronym{so}{SO}{Service Orchestrator}
\newacronym{lo}{LO}{Lifecycle Orchestrator}

\newacronym{tosca}{TOSCA}{Topology and Orchestration Specification for Cloud Applications}

\newacronym{onf}{ONF}{Open Networking Foundation}

\newacronym{sdn}{SDN}{Software Defined Networking}

\newacronym[longplural=Open Flow Switches,shortplural=OFSes]{ofs}{OFS}{Open Flow Switch}
\newacronym[longplural=OpenvSwitches,shortplural=OVSes]{ovs}{OVS}{OpenvSwitch}

\newacronym{fe}{FE}{Forwarding Element}
\newacronym{ne}{NE}{Network Element}
\newacronym{nos}{NOS}{Network Operating System}

\newacronym[longplural=Southbound Interfaces,shortplural=SBIs]{sbi}{SBI}{Southbound Interface}
\newacronym[longplural=Northbound Interfaces,shortplural=NBIs]{nbi}{SBI}{Northbound Interface}

\newacronym{onos}{ONOS}{Open Network Operating System}
\newacronym{e2e}{E2E}{End-to-End}

\newacronym{ns}{NS}{Network Service}

\newacronym{nfc}{NFC}{Network Function Chaining}

\newacronym{sfc}{SFC}{Service Function Chaining}

\newacronym{sf}{SF}{Service Function}

\newacronym{vcpe}{vCPE}{virtual Customer Premises Equipment}

\newacronym{nffg}{NF-FG}{Network Function Forwarding Graph}

\newacronym{nfib}{NF-IB}{Network Function Information Base}

\newacronym{un}{UN}{Universal Node}

\newacronym{sg}{SG}{Service Graph\glsadd{sgg}}

\newacronym{rg}{RG}{Resource Graph\glsadd{rgg}}

\newacronym{sm}{SM}{Service Management}
\newacronym{af}{AF}{Adaptation Function}

\newacronym{nsor}{NSO}{Network Service Orchestrator}

\newacronym{ro}{RO}{Resource Orchestrator}

\newacronym{ca}{CA}{Controller Adapter}

\newacronym{sl}{SL}{Service Layer}
\newacronym{ol}{OL}{Orchestration Layer}
\newacronym{cas}{CAS}{Controller Adaptation Sublayer}
\newacronym{ros}{ROS}{Resource Orchestration Sublayer}
\newacronym{il}{IL}{Infrastructure Layer}
\newacronym{mp}{MP}{Management Plane}
\newacronym{nfs}{NFS}{Network Functions System}

\newacronym{dov}{DoV}{Domain Virtualizer}
\newacronym{raf}{CA-RAF}{Controller Adapter-Resource Abstraction Function}
\newacronym{drdb}{DRDB}{Domain Resource Database}
\newacronym{pef}{PEf}{Policy Enforcement}
\newacronym{vcm}{VCM}{Virtual Context Manager}

\newacronym{sap}{SAP}{Service Access Point}

\newacronym{osm}{OSM}{Open Source MANO}
\newacronym{openo}{Open-O}{Open-Orchestrator}
\newacronym{5gex}{5G-Ex}{5G-Exchange}
\newacronym{aria}{ARIA}{Agile Reference Implementation of Automation}
\newacronym{onap}{ONAP}{Open Network Automation Platform}
\newacronym{5gppp}{5G-PPP}{5G Infrastructure Public Private Partnership}
\newacronym{sonata}{SONATA}{Service Programming and Orchestration for Virtualized Software Networks}

\newacronym{lso}{LSO}{Lifecycle Service Orchestration}

\newacronym{dc}{DC}{Data Center}
\newacronym{cn}{CN}{Compute Node}

\newacronym[longplural=Big Switch and
Big Software Nodes,shortplural=BiS-BiS Nodes]{bsbs}{BiS-BiS}{Big
  Switch with Big Software}

\newacronym{spdevops}{SP-DevOps}{Service Provider DevOps}

\newacronym{op}{OP}{Observability Point}

\newacronym{mf}{MF}{Monitoring Function}

\newacronym{rpc}{RPC}{Remote Procedure Call}
\newacronym{rest}{REST}{REpresentational State Transfer}

\newacronym[longplural=Intrusion Detection Systems]{ids}{IDS}{Intrusion Detection System}

\newacronym{idsc}{IDSC}{\gls{ids} Control}
\newacronym{eids}{E-IDS}{Elastic \gls{ids}}
\newacronym{eidsc}{E-IDSC}{Elastic \gls{idsc}}

\newacronym{vbaas}{VBaaS}{\gls{vnf} Benchmarking as a Service}

\newacronym{ib}{IB}{Information Base}

\newacronym{actn}{ACTN}{Abstraction and Control of Transport Networks}
\newacronym{opnfv}{OPNFV}{Open Platform for NFV}
\newacronym{ims}{IMS}{IP Multimedia Subsystem}
\newacronym{cdn}{CDN}{Content Delivery Network}

\newacronym{sdwan}{SD-WAN}{Software Defined Wide Area Network}

\newacronym{ngmn}{NGMN}{Next Generation Mobile Networks}

\newacronym{lte}{LTE}{Long Term Evolution}

\newacronym{oasis}{OASIS}{Organization for the Advancement of Structured Information Standards}

\newacronym{pce}{PCE}{Path Computation Element}
\newacronym{mpls}{MPLS}{Multi-Protocol Label Switching}
\newacronym{lsp}{LSP}{Label Switching Path}

\newacronym{wg}{WG}{Working Group}

\newacronym{nfvrg}{NFVRG}{\gls{nfv} Research Group}
\newacronym{sdnrg}{SDNRG}{\gls{sdn} Research Group}

\journal{Computer Communications}

\begin{document}

\begin{frontmatter}



\title{Network Service Orchestration: A Survey}


\author[First]{Nathan F. Saraiva de Sousa}
\author[First]{Danny A. Lachos Perez}
\author[First]{Raphael V. Rosa}
\author[Second]{Mateus A. S. Santos}
\author[First]{Christian Esteve Rothenberg}

\address[First]{Department of Computer Engineering and Industrial Automation, School of Electrical and Computer Engineering, University of Campinas - UNICAMP, Campinas, SP, Brazil}
\address[Second]{Ericsson Research, Indaiatuba, SP, Brazil}

\fntext[*]{Any feedback is welcome to improve the work turning the \textit{github} and \textit{arxiv} versions of this publication a ``living document'' driven by community contributions as NSO evolves.  Do not hesitate to contact the authors and submit \textit{github} pull requests or issues: 
https://github.com/intrig-unicamp/publications/tree/master/NSO-Survey.}

\begin{abstract}
Business models of network service providers are undergoing an evolving transformation fueled by vertical customer demands and technological advances such as 5G, Software Defined Networking~(SDN), and Network Function Virtualization~(NFV). Emerging scenarios call for agile network services consuming network, storage, and compute resources across heterogeneous infrastructures and administrative domains. Coordinating resource control and service creation across interconnected domains and diverse technologies becomes a grand  challenge. Research and development efforts are being devoted to enabling orchestration processes to  automate, coordinate, and manage the deployment and operation of network services. In this survey, we delve into the topic of Network Service Orchestration~(NSO) by reviewing the historical background, relevant research projects, enabling technologies, and standardization activities. We define key concepts and propose a taxonomy of NSO approaches and solutions to pave the way to a common understanding of the various ongoing efforts towards the realization of diverse NSO application scenarios. Based on the analysis of the state of affairs, we present a series of open challenges and research opportunities, altogether contributing to a timely and comprehensive survey on the vibrant and strategic topic of network service orchestration.

\end{abstract}


\end{frontmatter}


\section{Introduction}

Telecommunication  infrastructures consist of a myriad of technologies from specialized domains such as radio,  access, transport, core and (virtualized) data center networks. Designing, deploying and operating end-to-end services are commonly   manual and long processes performed via traditional \gls{oss} resulting in long lead times (weeks or months) until effective service delivery~\cite{BluePlanet2017ProductsOrchestration}. Moreover, the involved workflows are commonly hampered by built-in hazards of infrastructures strongly coupled to physical topologies and hardware-specific constraints.

\begin{figure}[t!]
  \centering
  \includegraphics[scale=.76]{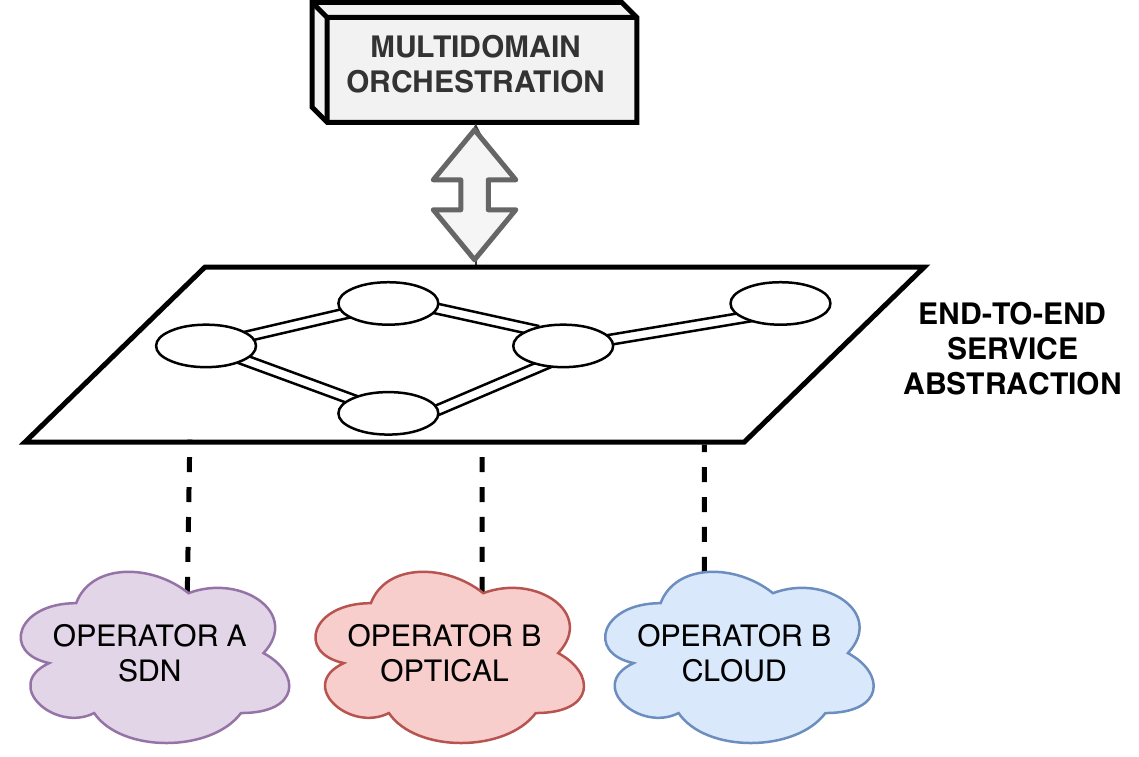}
    \caption{Context and scope of Network Service Orchestration.}
    \label{intro}
\end{figure}

Technological advances under the flags of \gls{sdn} \cite{surveySDN} and \gls{nfv} \cite{Mijumbi2016NetworkChallenges} bring new ways in which network operators can create, deploy, and manage their services. \gls{sdn} and \gls{nfv}, as well as cloud computing introduce new means for efficient and flexible utilization of their infrastructures through a software-centric service paradigm \cite{Sonkoly2014UNIFYingView}. However, to realize this paradigm, there is a need to model the end-to-end service and have the ability to abstract and automate the control of physical and virtual resources delivering the service. The coordinated set of activities behind such process is commonly referred to as \textit{orchestration}. In general, orchestration refers to the idea of automatically selecting and controlling multiple resources, services, and systems to meet certain objectives (e.g., a customer requesting a specific network service). Altogether, the process shall be timely, consistent, secure, and lead to cost reduction due to automation and virtualization. We refer to \gls{nso} as the automated management and control processes involved in end-to-end services deployment and operations  performed mainly by telecommunication operators and service providers, involving different types of resources and potentially multiple operators, as illustrated in Figure~\ref{intro}. 

\begin{figure}[t!]
  \centering
  \includegraphics[scale=.25]{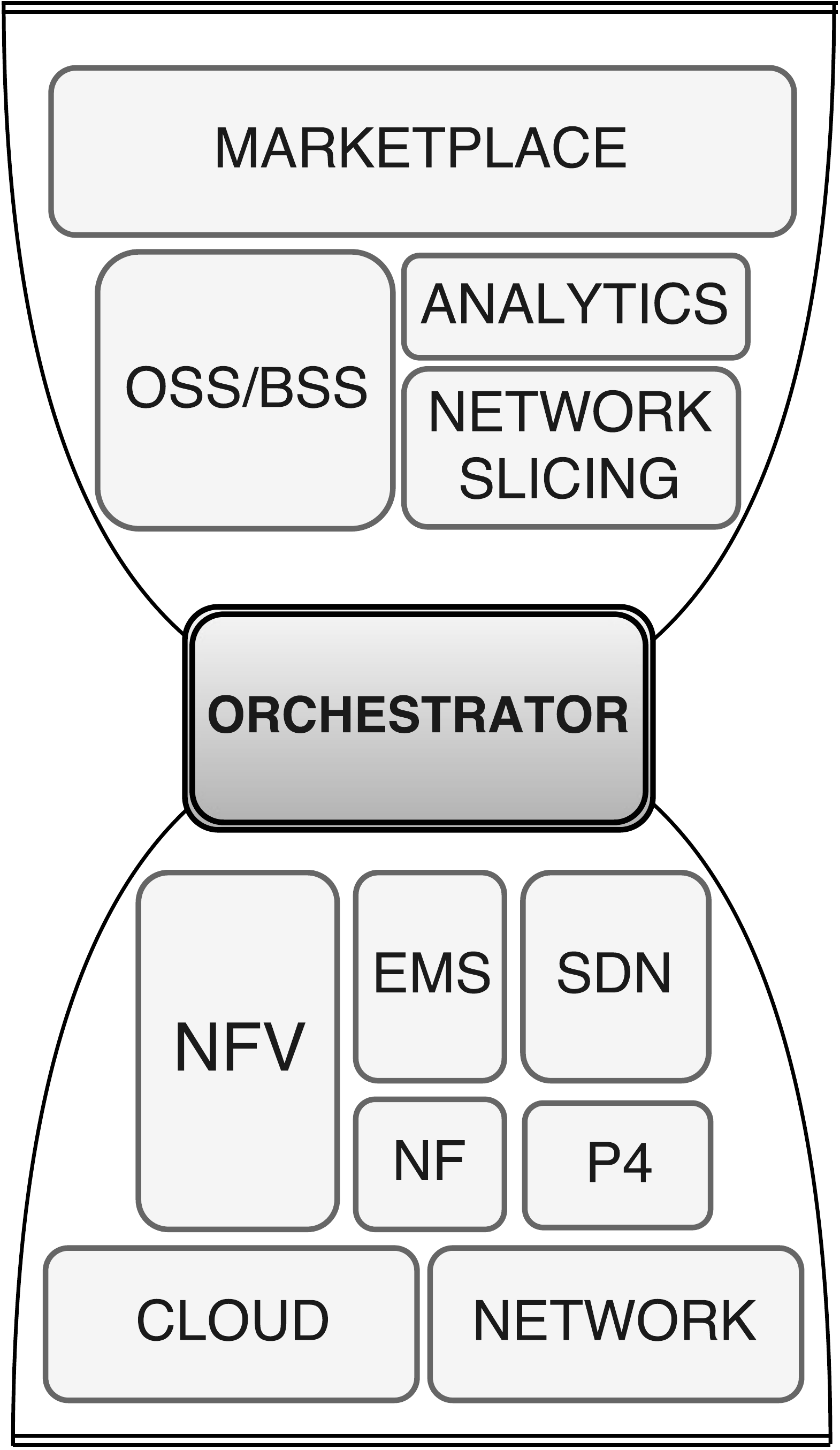}
    \caption{Strategic role of the \protect\gls{nso} as the glue between the actual services and the underlying management of resources.}
    \label{orch}
\end{figure}

Dealing with the holistic nature of network services, \gls{nso} is responsible for decoupling the high-level service layer (e.g., applications, service  slices, \gls{oss}) from the underlying management and resources layers (e.g., controllers, \gls{ems}, \gls{vim}), providing agility, enabling innovative service, optimizing resources, and altogether delivering a more flexible infrastructure for tailored services delivery. By introducing service abstractions through well-defined data models, descriptors, and programming interfaces, \gls{nso} defines the interaction with (chains of) network functions in underlying technologies and infrastructures through a unifying pane glass for service definition, deployment, and operation. For example, NSO may connect traditional OSS/BSS to network functions running in virtualized infrastructures. This unifying tenet is illustrated through  the hourglass shape in Figure~\ref{orch}, where the  significance of NSO is highlighted as an inter-working technology-agnostic glue --- drawing an analogy to IP in the traditional network protocol stack. 

As today, broad understanding and practical definitions of \gls{nso} are still missing --- not only across but also inside networking communities. The maturity of ongoing efforts varies largely with the overall technical approach being very much fragmented and showing little consolidation around an overarching notion of network service orchestration. 

The main objective of this survey is to provide a comprehensive understanding of the research, standardization, and software development efforts around the overcharged term of  \acrlong{nso}. We present an in-depth and up-to-date study on network service orchestration covering some historical background and context, enabling technologies, standardization activities, actual solutions, open challenges, and research opportunities. We propose a taxonomy of the main characteristics and features of NSO approaches. We also make the mapping of the \gls{nso} primary characteristics and technical implementations to current open source platforms and research projects.    

Throughout the survey, we distinguish between two types of domains. First, \textit{administrative domains}, which map to different organizations and therefore may exist within a single service provider or cover a set of service providers. In one administrative domain, multiple \textit{technology domains} can exist based on the type of technology in scope, for example, Cloud, \gls{sdn}, \gls{nfv}, or Legacy. 
Broadly speaking, we refer to \gls{nso} as the automated coordination of resources and services embracing both single-domain and multi-domain footprints.  

\begin{figure*}[th]
  \centering
  \includegraphics[scale=.5]{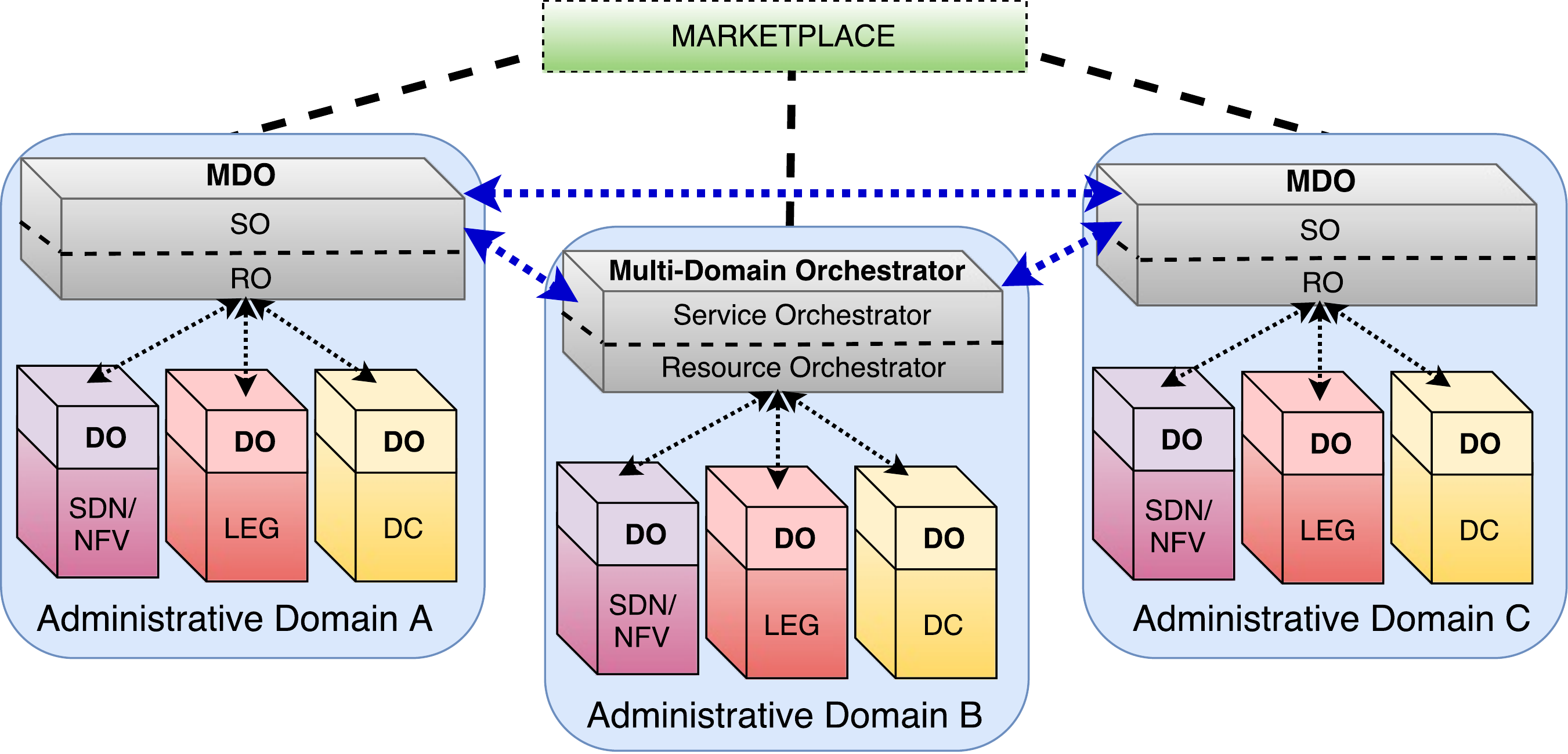}
    \caption{High-level reference model to illustrate the scope of \protect\acrfull{nso} in single-domain and multi-domain environment. The \protect\gls{nso}  need to have an overview of entire environment to compose the service mainly if it uses resources of different domains.}
    \label{mdo}
\end{figure*}

Figure~\ref{mdo} presents a generic high-level reference model for multi-domain Network Service Orchestration, featuring a \gls{mdo} per administrative realm and including the notion of a Marketplace for business interactions. 
\glspl{mdo} coordinate resources and services in a multiple administrative domain scope covering multiple technology domains~\cite{5GPPPArchitectureWorkingGroup2016ViewArchitecture}. 
The exchange of information, resources, and services themselves are essential components of an end-to-end network service delivery.  The \gls{mdo} exposes the available services to the marketplace allowing service providers to sell network services directly to their customers or other providers under various possible resources consumption models (e.g., trading resources from each other). 
The \gls{mdo} can be seen as a single element with a possible split into two functional components: \gls{so} and \gls{ro}. The \gls{so} orchestrates high-level services while the \gls{ro} is responsible for managing resource and orchestrating workflows across technology domains. 
The \glspl{do} perform orchestration in each local domain acting on the underlying infrastructures and exposing resources and network functions northbound to the \gls{mdo}. 

\noindent \textbf{Related work.} Several works have somehow addressed the theme of orchestration in different scopes, including clouding computing \cite{Weerasiri2017}, \gls{sdn}~\cite{Jarraya2014},~\cite{surveySDN}, and \gls{nfv}~\cite{YongLi2015Software-DefinedSurvey},~\cite{Mijumbi2016NetworkChallenges},~\cite{Bhamare2016}. In~\cite{Weerasiri2017}, for example, the authors propose a taxonomy and survey of cloud orchestration techniques. However, its scope is limited to cloud resources. Saadon et al.~\cite{SAADON201917} presented an overview of orchestration standardization efforts and implementation in next-generation network management. However, while our approach has reviewed the broad NSO scope in Standards Developing Organizations (SDOs), Saadon et al. only specified the interactions of the orchestration layer and the OSS application layer. Vaquero et al.~\cite{VAQUERO201920} introduced next-generation orchestration techniques with a focus on the leading challenges  involved in diverse end-to-end orchestration processes. Differently, our survey  embraces the solution space by dissecting multi-faceted orchestration aspects and approaches from different communities.  

The work of Rotsos~et~al.~\cite{Rotsos2017NetworkSurvey} is arguably the first notable attempt to survey the realm of network service orchestration. The authors provide an analysis of diverse standardization activities around \gls{nso} from an operator perspective. The article follows a top-down approach, defining terminologies, requirements, and objectives of a network service orchestrator. 
In contrast, our definition and approach to \gls{nso} are distinct. We follow a systems-oriented and broadly generic approach,  where \gls{nso} encompasses high-level services as defined by telecommunications operators along business and technological operations for network service instantiation and run-time operation. Besides, we condense standardization efforts, challenges, solutions, and research projects of NSO landscape into single work. Most significantly, we feature 150+ references providing a broader scope covering:
\begin{itemize}
\item Historical overview of the very much overloaded orchestration term;
\item Comprehensive approach to NSO by clarifying fundamental aspects of core NSO functions, and how different research and standardization are addressing the topic;
\item Taxonomy of the main aspects of any NSO solution;
\item Up-to-date review of related standards activities;
\item Overview of relevant research projects and software frameworks.
\end{itemize}

\begin{figure*}[t!]
  \centering
  \includegraphics[scale=.62]{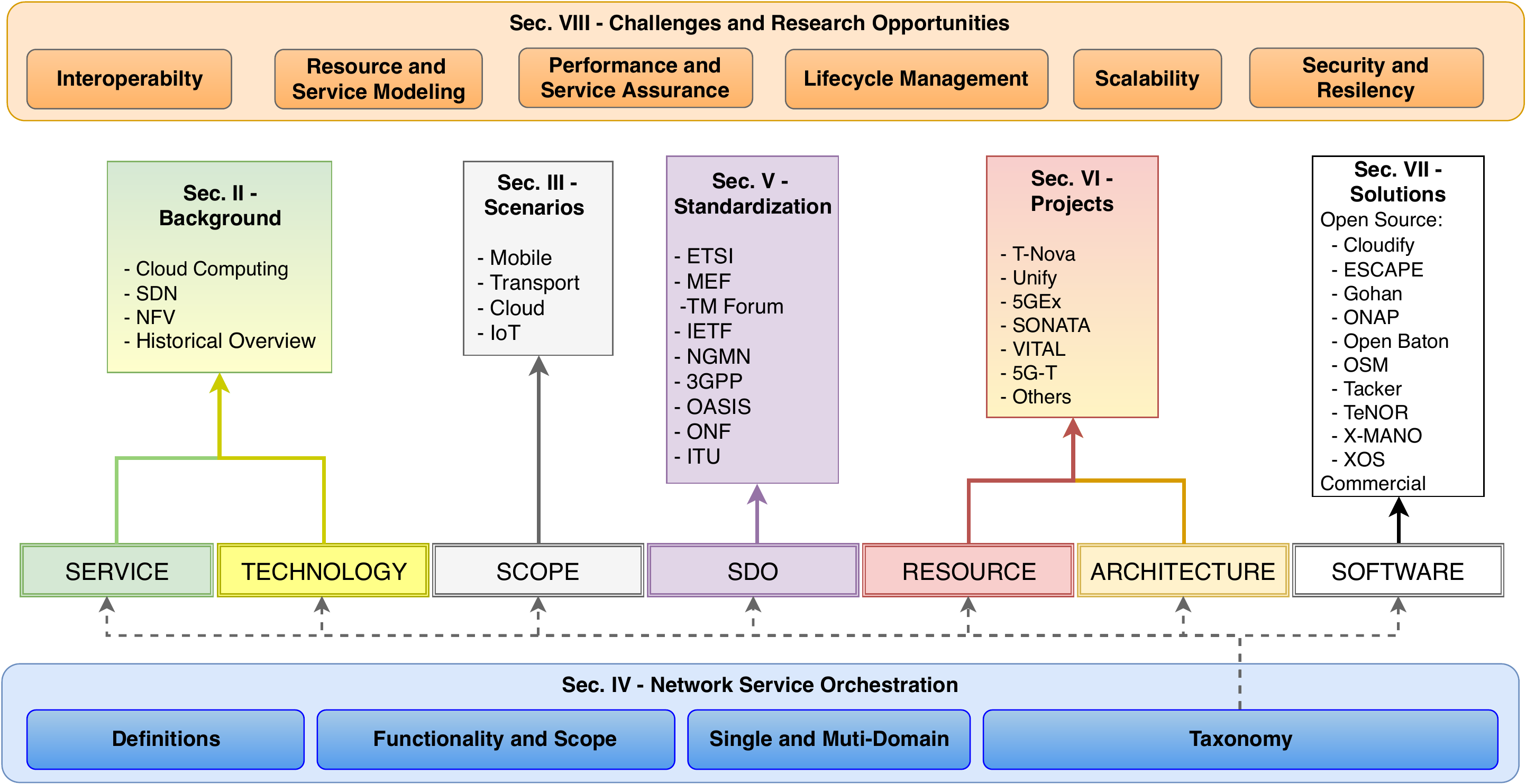}
    \caption{Overview of the organization of this survey on NSO.}
    \label{org}
\end{figure*}

\noindent \textbf{Survey Organization.}  The survey is organized as depicted in Figure~\ref{org}. Section~\ref{sec:background} presents essential background and key technologies related to network service orchestration: Cloud computing, \gls{sdn}, \gls{nfv}, historical overview of orchestration, and the relationship between all mentioned technologies. Section~\ref{sec:scneario} outlines four potential scenarios to illustrate the \gls{nso} in practice. Concepts, functions, scope, and an NSO taxonomy split into seven key aspects are presented in Section~\ref{sec:nso}. Section~\ref{sec:stand} focuses on the standardization outcomes produced by nine important organizations,   whereas Section~\ref{sec:project} covers six major research projects around \gls{nso}. Section~\ref{sec:proj} provides an overview of ten open source solutions and some commercial initiatives.  The discussion in Section~\ref{sec:challenge} points to six groups of open challenges and research opportunities. Finally, Section~\ref{sec:Conclusion} concludes the survey.
\section{Background}
\label{sec:background}

\gls{nso} foundations can be rooted back to three enabling technologies, namely Cloud Computing, SDN, and NFV. This section provides a brief background on these topics and their relationships to NSO, in addition to a short historical review of the term ``orchestration''.

\subsection{Cloud Computing}
Cloud computing is a model for providing resource virtualization (e.g., networks, servers, storage, and services) with high flexibility, cost efficiency, and centralized management~\cite{Le2016SurveyNetworks}. The cloud computing service models are generally categorized in \gls{iaas}, \gls{paas}, and \gls{saas} which offer, respectively,  virtual resources (compute, storage, and network), software and development platforms (provided by the cloud infrastructure), and Internet-based applications (hosted on the cloud)~\cite{bele2018empirical}.

In a cloud environment, the notion of orchestration has also been used for integrating basic services~\cite{Vouk2008CloudImplementations}. Orchestration in the cloud involves dynamically deploying, managing and maintaining resource and services across multiple heterogeneous cloud platforms in order to meet the needs of clients. 

\subsection{Software Defined Networking (SDN)}
\gls{sdn}~\cite{surveySDN} is an evolving networking paradigm that attempts to resolve the strongly vertical integration of current network environments. To this end, \gls{sdn} proposals decouple the control plane (i.e., control logic) from the data plane (i.e., data forwarding equipment). With this new architecture, routers and switches become simple forwarding network elements whose control logic is provided by a logically centralized external entity called \gls{sdn} controller or \gls{nos}.

In multi-domain scenarios, there are different \gls{sdn} controllers deployed to manage specific segments of a network (e.g., fronthaul, backhaul, and core). \gls{sdn} implements some level of resource orchestration in order to coordinate control plane actions with multiple \gls{sdn} controllers. By recognizing the needs of higher-level  orchestrator(s), \gls{sdn} controllers can be programmed to monitor the network and make automated (real time) decisions in case of security problems, faulty devices, traffic congestion, among others~\cite{SDXCentral2015WhatSDNOrc}. 

\subsection{Network Function Virtualization (NFV)}
\label{subsec:nfv}

According to \gls{etsi} \gls{isg} \gls{nfv} \cite{ETSIIndustrySpecificationGroupISGNFV2014NetworkNFV},  \acrlong{nfv} is responsible for separating network functions from the hardware and offering them through virtualized services, decomposed into \glspl{vnf}, on general purpose servers. With the virtualization of the network functions, \gls{nfv} promises more flexible and faster network function deployment, as well as dynamic scaling of the \glspl{vnf} towards providing finer settings. In \gls{nfv} environment, new services do not require new hardware infrastructure, but simply the software installation, i.e., to create \glspl{vnf}.

\begin{figure}[t]
  \centering
  \includegraphics[scale=.46]{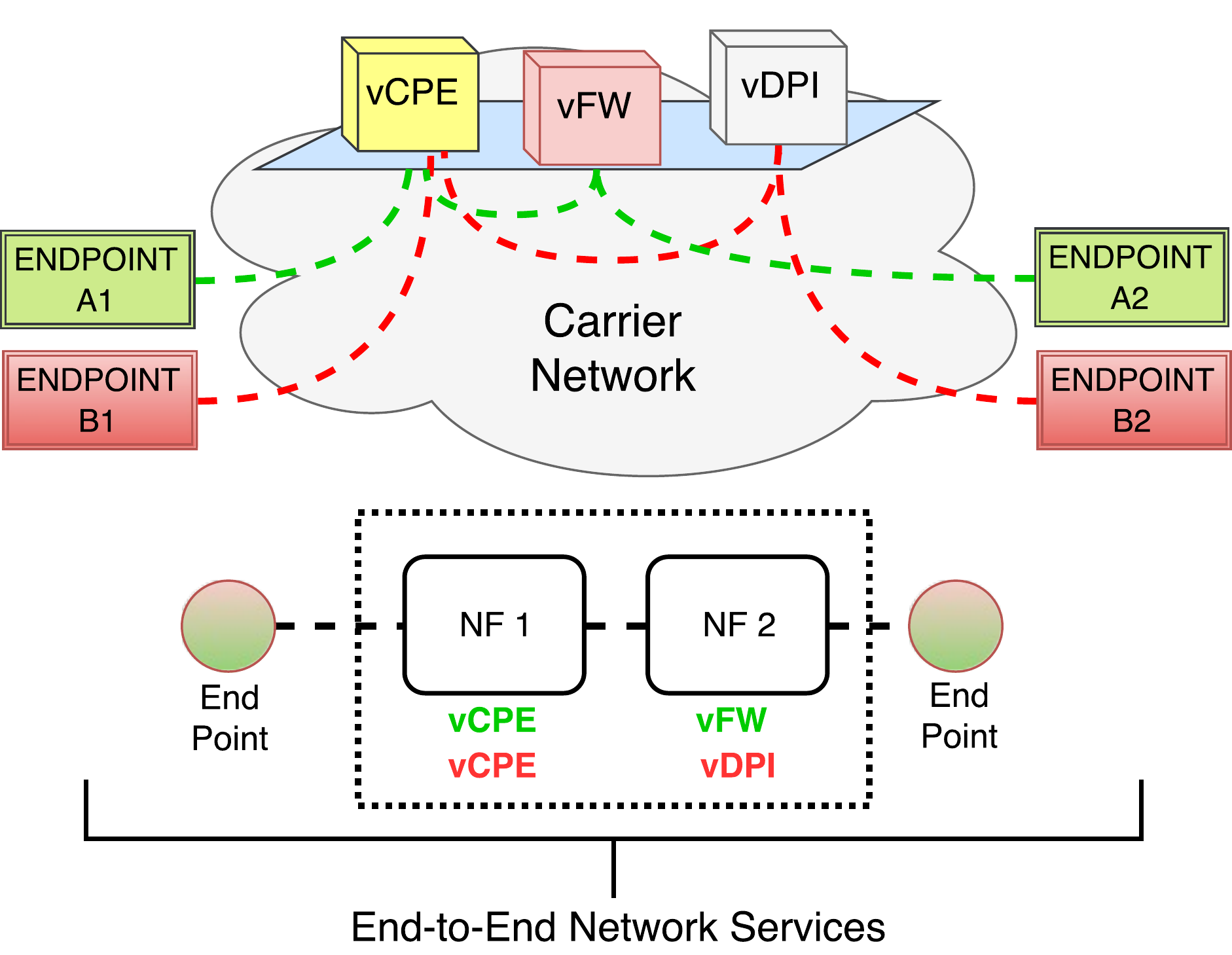}
    \caption{Example of two end-to-end network services composed of two \protect\glspl{nf} each. NFV enables the reuse of \protect\glspl{vnf}, e.g., \protect\gls{vcpe}.}
    \label{nffg}
\end{figure}

\glspl{vnf} can be connected or combined as building blocks to offer a full-scale network communication service. This connection is known as service chain. Within the scope of the \gls{isg} \gls{nfv} \cite{ETSIIndustrySpecificationGroupISGNFV2014NetworkNFV}, service chain is defined as a graph of logical links connecting \glspl{nf} towards describing traffic flow between these network functions. This is equivalent to the \gls{sfc}~\cite{Halpern2015} defined by Service Function Chaining Working Group (IETF SFC WG) of the \gls{ietf}.  
An end-to-end network service may cover one or more \gls{nffg} which interconnect \glspl{nf} and end points.  Figure~\ref{nffg} describes two examples of end-to-end network services. The first (green line) is composed of \gls{vcpe} and \gls{vfw} \glspl{vnf} and two endpoints (A1 and A2). The second (red line) is composed of \gls{vcpe} and \gls{vdpi} \glspl{vnf} and two endpoints (B1 and B2). In the examples, a single VNF can be part of one or more network services. It emphasizes the multi-tenant aspect of \gls{nfv}.

\begin{figure}[t]
  \centering
  \includegraphics[scale=.45]{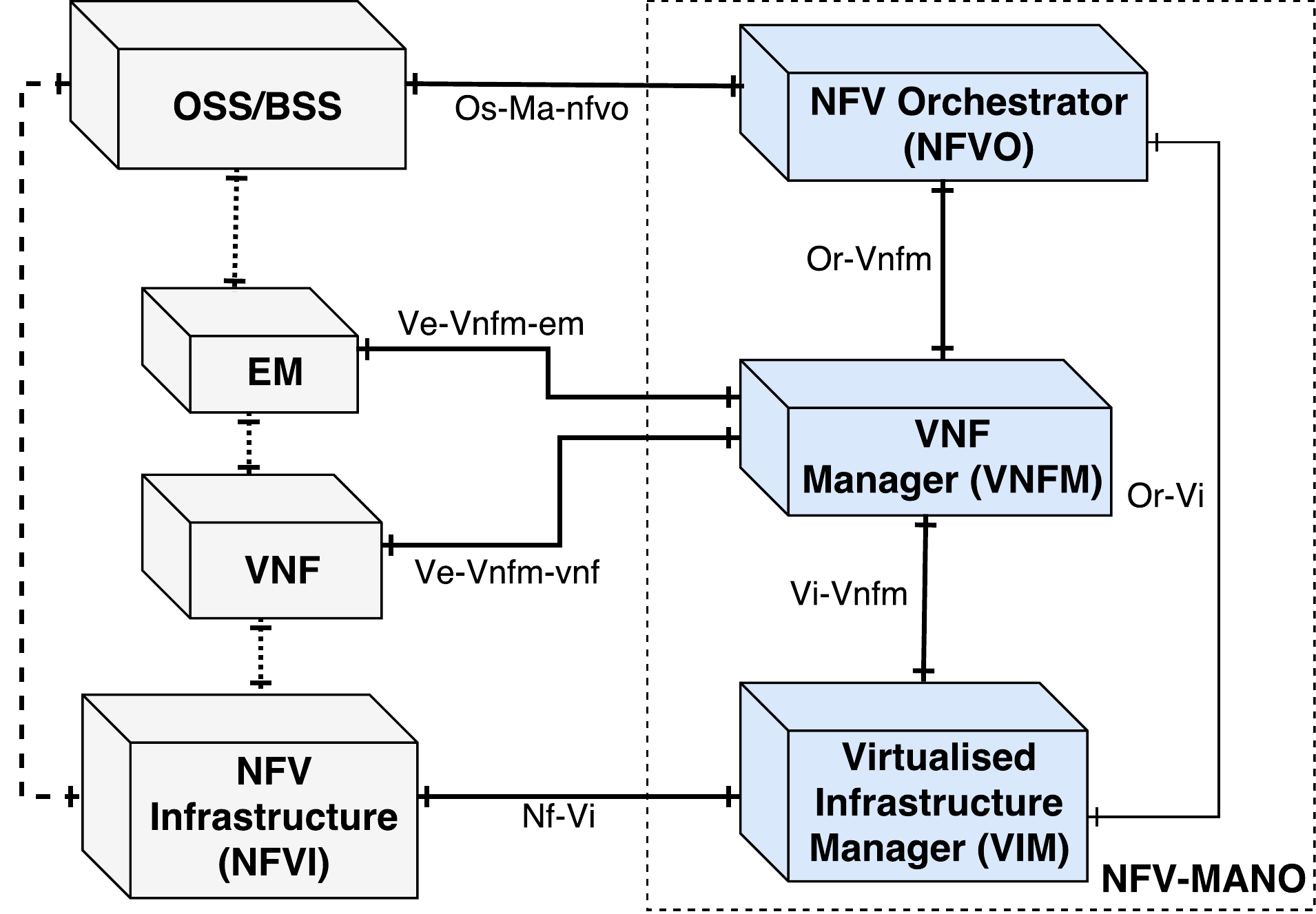}
    \caption{The \protect\gls{nfvmano} architectural framework. Adapted from \protect\cite{ETSIIndustrySpecificationGroupISGNFV2014NetworkOptions}}
    \label{mano}
\end{figure}

\gls{etsi} has developed a reference architectural framework and specifications in support of NFV management and orchestration. The framework covers orchestration and lifecycle management of physical and virtual resources. According to~\cite{ETSIIndustrySpecificationGroupISGNFV2013NetworkFramework}, ``the framework is described at a functional level and it does not propose any specific implementation." Figure~\ref{mano} shows the \gls{etsi} \gls{nfv}-\acrfull{mano} architectural framework with their main functional blocks~\cite{ETSIIndustrySpecificationGroupISGNFV2014NetworkOptions}:

 \noindent \textbf{Operation/ Business Support System (OSS/BSS)}: block responsible for operation and business applications that network service providers use to provision and operate their network services. It is not tightly integrated into the \gls{nfv}-\gls{mano} architecture.

\noindent \textbf{\gls{em}}: component responsible for the network management functions FCAPS (Fault, Configuration, Accounting, Performance, and Security) of a running \gls{vnf}.

\noindent\textbf{\gls{vnf}}: functional block representing the Virtualized Network Function implemented on a physical server. For instance, Router \gls{vnf}, Switch \gls{vnf}, Firewall etc.

\noindent \textbf{\gls{nfvi}}: representing all the hardware (compute, storage, and networking) and software components where \glspl{vnf} are deployed, managed and executed. 

\noindent \textbf{\gls{nfvo}}: it is the primary component, in charge of the orchestration of \gls{nfvi} resources across multiple \glspl{vim} and lifecycle management of network services. 

\noindent\textbf{\gls{vnfm}}: performs configuration and \gls{vnf} lifecycle management (e.g., instantiation, update, query, scaling, termination) on its domain.

\noindent \textbf{\gls{vim}}: block that provides controlling and managing the \gls{nfvi} resources as well the interaction of a \gls{vnf} with hardware resources. For example, OpenStack as cloud platform and OpenDaylight and \gls{onos} as \gls{sdn} controllers.

In the \gls{nfv} context, \gls{etsi} \gls{nfvmano} defines the orchestrator with two main functions including \textit{resources orchestration across multiple \glspl{vim}} and \textit{network service orchestration}~\cite{GSNFV-MAN001:2014}. Network service orchestration functions provided by the \gls{nfvo} are listed below:
\begin{itemize}
\item Management of Network Services templates and \gls{vnf} Packages. 
\item Network Service instantiation and management;
\item Management of the instantiation of \glspl{vnfm} and \glspl{vnf} (with support of \glspl{vnfm});
\item Validation and authorization of \gls{nfvi} resource requests from \gls{vnf} managers;
\item Policy management related to affinity, scaling (auto or manual), fault tolerance, performance, and topology.
\end{itemize}

\gls{etsi} \gls{nfvo} functions regarding Resource Orchestration include: (\textit{i})  Orchestration of NFVI resources across multiple \glspl{vim}, (\textit{ii})  \gls{nfvi} resource management including compute, storage and network, and (\textit{iii}) collect usage information of \gls{nfvi} resources. \gls{nfvo} functions defined by \gls{etsi} are limited to the delivery of network services, i.e., without being aware of what type of service has been instantiated.

The \gls{nfvmano} reference architecture is not  specific about \gls{sdn} in its architecture but  assumes that necessary transport infrastructure is already established and ready to be used. However, work at \gls{etsi} identifies use cases and the most common options for using SDN in an NFV architectural framework~\cite{ETSINetworkFramework}. The document also points to  proof of concepts and recommendations towards such integration work.
\cite{nfv-survey18} provides a recent in-depth survey on NFV state of affairs. 

\subsection{Orchestration: Historical Overview}
The academic community and industry generally require some time to define the real meaning, reach and context of the concepts related to new technology trends as is the case with the term \textit{Orchestration}. 
The term orchestration is used in many different areas, such as multimedia, music, \gls{soa}, business processes, Cloud, \gls{sdn}, and, more recently, in \gls{nfv}.

From an end-user perspective, orchestration reminds a symphony orchestra where a set of instruments play together according to an arrangement. The music is arranged and split into small parts, after assigns to different musical instruments. When, who, and what will be played, as well as the conducting are essential parts towards achieving the desired effect. In next paragraphs, we identified the first works that use the term orchestration in other areas. 

One of the first works in the \gls{ict} area that cites the term orchestration is \cite{Anderson1983} in 1983. It discusses that an autonomous system will require orchestration of the behavior of the entire system in order to obtain autonomy, interdependence and artificial intelligence. The authors in~\cite{Campbell1992} relate orchestration with the coordination and control of multiple media traffics. It distinguishes orchestration from synchronization and defines an architecture where orchestration acts in different layers. In the same scope, \cite{Robbins1997ImplementationArchitecture} relates the term to multimedia data, where orchestration is associated with multimedia presentation lifecycle management involving the coordination of stages that constitute all orchestration processes. 

The use of orchestration is also widely discussed in the scope of web services. In this context, orchestration and automation are considered separate processes. The work in~\cite{Peltz2003WebChoreography} defines orchestration like an executable process that can interact both internal and external services and must be dynamic, flexible, and adaptable to changes. It emphasizes that orchestration describe how web services can act with each other at the message level, including the business logic and execution order of the activities. 

The authors in~\cite{Grit2006} present the term orchestration in the context of virtual resource management. They define orchestration as a process that involves all the necessary steps to map the application (running on a virtual machine) onto shared underlying infrastructure. 

Orchestration in the cloud environment is well-known and refers to locating, coordinating and selecting resources, including compute, storage and virtual networks to fulfill the desired requirements. The authors in \cite{Galis2009ManagementInternet} provide an overview of networking architecture definition for the \gls{fi} based on the concepts of cloud computing. One of the pillars for the \gls{fi} pointed out by the article is Orchestration. In the envisioned architecture, the orchestration function is to coordinate the integrated behavior and operations to dynamically adapt and optimize resources in response to changing context following business objectives and policies.

In the \gls{sdn} landscape, orchestration refers to an overarching function to manage and automate the network behavior~\cite{5984813}. More recently in 2012~\cite{ETSI2012NetworkAction}, orchestration has been generally related to \gls{nfv} environments mainly through its reference architecture and its \gls{nfv} Orchestrator component (more details in Subsection~\ref{subsec:nfv}).  

Currently, the scope of orchestration has become broader and encompasses automation of the end-to-end network service lifecycle. According to \cite{MEF:Third:2015}, service orchestration refers to the programmatic control of underlying infrastructure including existing networks and enabling technologies, such as SDN and NFV.

From the existing and evolving definitions around orchestration presented, we can derive certain relationships between orchestration, automation, and management. Although the three terms are often lumped together, it is necessary an understanding of the differences between them as they are not the same thing. Automation describes a simple and technical task without the human intervention, for example, launching a web server, stopping a server. Management is responsible for maintaining and healthiness of infrastructure. Its role consists of activities such as alarms for event detection, monitoring, backups of critical systems, upgrades, and license management. Orchestration, in turn, is concerned with the execution of a workflow (processes) in the correct order. It controls the overall workflow process from starting the service until it ends with the objective to optimize and automate the network service deployment. 

\begin{figure}[t]
    \centering
    \includegraphics[scale=.45]{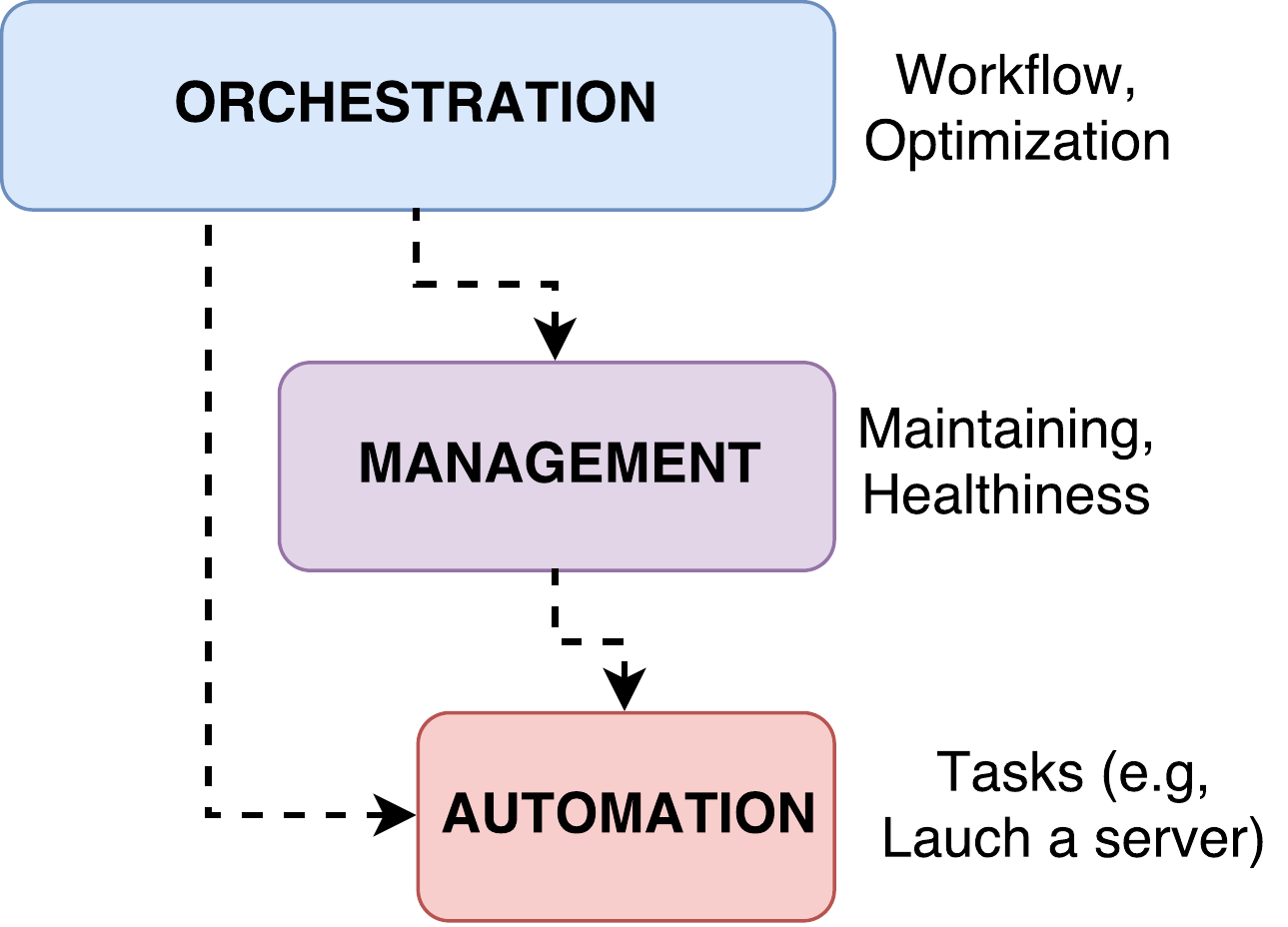}
      \caption{Relationship among orchestration, management, and automation. Both orchestration and management use automation in their processes.}
      \label{diff}
\end{figure}

Figure~\ref{diff} illustrates the relationship among orchestration, management, and automation through a hierarchy. Orchestration is a high-level plane followed by  the management layer. Automation lies at the lower layer. In our vision, the orchestration layer depends on tasks performed by management. Both management and orchestration are based on the use of automation in the execution of their tasks. Nevertheless, several activities are only performed by a certain function: optimization, for instance, cannot be achieved through simple automation. 

\begin{figure}[t]
  \centering
  \includegraphics[scale=.58]{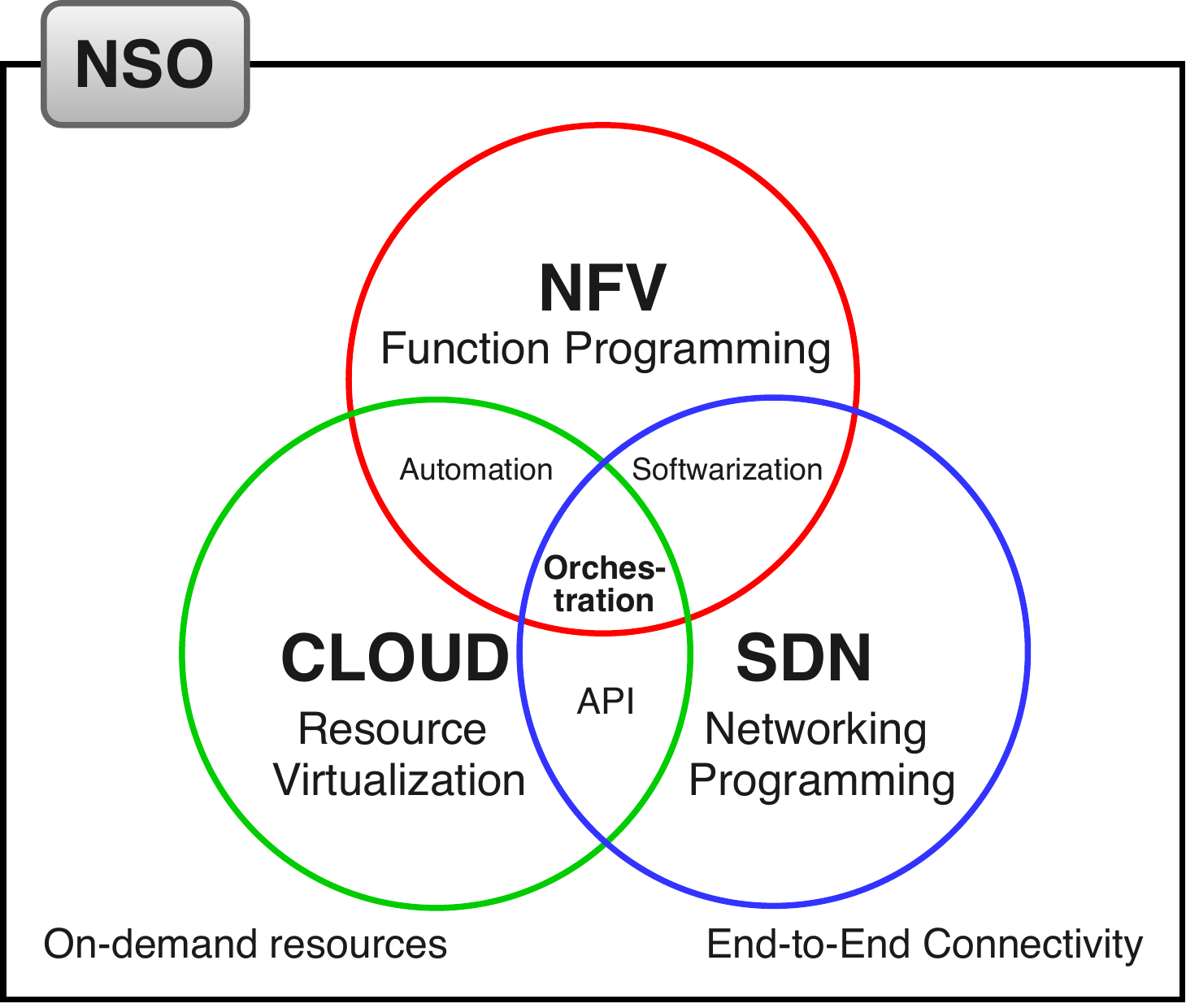}
    \caption{Illustration of relationships among \protect\gls{nso}, \protect\gls{nfv}, \protect\gls{sdn}, and Cloud.}
    \label{nso_rel}
\end{figure}

Based on all the above-mentioned background, in short, \gls{nso} is in charge of the full network service lifecycle to deliver end-to-end connectivity along additional services. To this end, orchestration is supported by advances in cloud computing, and technologies such as \gls{sdn} and \gls{nfv}, which offer the ability to reconfigure the network quickly as well as programming the forwarding and processing of the traffic. Figure~\ref{nso_rel} shows how \gls{nso}, \gls{nfv}, \gls{sdn}, and Cloud Computing work together.

Each one of these paradigms/technologies has different functions: high level orchestration for \gls{nso}, function programming for \gls{nfv}, networking programming for \gls{sdn},  and resource virtualization for cloud computing. Note that such technologies are complementary in order to provide complete management of the network services lifecycle. Although they have different functions, they share a common feature: \textit{orchestration}. They can work in an integrated pattern to offer advantages such as agility, cost reduction, automation, softwarization, and end-to-end connectivity, to enable novel services and applications such as 5G networks.

Our goal in this subsection was to set the ground and identify the main areas in which the term orchestration is inserted and how it is approached at a high level. An overview of the term usage is illustrated in the timeline of Table~\ref{timeline}. The focus of this survey is to detail the orchestration process in the context of the implementation and operation of network services by operators and service providers.

\begin{table}[t!]
\small
\caption{Historical timeline of term orchestration }\vskip -1ex
\label{timeline}
\begin{tabular}{@{\,}r <{\hskip 2pt} !{\foo} >{\raggedright\arraybackslash}p{5cm}}
\addlinespace[3ex]
\toprule
1983 & Autonomous system~\cite{Anderson1983}\\[13.5pt]
1992 & Media Traffic~\cite{Campbell1992}\\[7.5pt]
1997 & Multimedia presentation lifecycle management~\cite{Robbins1997ImplementationArchitecture}\\[1pt]
2003 & Web Service~\cite{Peltz2003WebChoreography}\\[4.5pt]
2006 & Virtual resource management~\cite{Grit2006}\\[4.5pt]
2009 & Cloud computing~\cite{Galis2009ManagementInternet}\\[3pt]
2011 & Software Defining Network~\cite{5984813}\\[1.5pt]
2012 & Network Function Virtualization~\cite{ETSI2012NetworkAction}\\[2pt]
2015 & Lifecycle Service Orchestration~\cite{MEF:Third:2015}\\
\end{tabular}
\end{table}
\section{Application Scenarios}
\label{sec:scneario}

\gls{nso} is envisaged to support diverse use case scenarios. This section aims at providing a brief practical view on a number of application domains and the main benefits provided by \gls{nso} in each scenario, delivering a sample of the expected potential of NSO in operation. 

\subsection{Next Generation Mobile Telecommunication Networks}

The fifth generation of mobile communication systems (5G) is expected to meet diverse and stringent requirements that are currently not supported by current mobile telecommunication networks, like ubiquitous connectivity (connectivity available anywhere), zero latency (lower than few milliseconds) and high-speed connection (10 times higher than 4G).

An efficient realization of 5G requires a flexible and programmable infrastructure covering transport, radio, and cloud resources~\cite{NGMN:5G:2017}. \gls{sdn} and \gls{nfv} are considered key enabling technologies to provide the required flexibility in processing and programmability, whereas end-to-end orchestration is regarded fundamental to improve the mobile service creation and resource utilization across all network segments, from radio access to transport~\cite{rostami-ran-transport-17}.

Furthermore, end-to-end orchestration should tackle  a significant challenge in mobile telecommunication networks, namely, the integration of different technologies, including radio, \gls{sdn} and \gls{nfv} so that network services may be dynamically created and adapted across the domains (wireless, aggregation and core). 

Finally, mobile management and orchestration solutions are expected to enable (i) congestion handling per subscriber or traffic, (ii) dynamic allocation of resources according to traffic variation and/or service requirements, and (iii) load reduction on transport networks and central processing units~\cite{EricssonInc.2015}. 

Future mobile/5G and fixed networks scenarios with diverse service requirements represent a growing and more complex challenge at the time of managing network resources. Network Slicing is being widely discussed in standard organizations as an essential mechanism to provide flexibility in the management of network resources~\cite{NGMN:5G:2017}. Network Slicing enables operators to create multiple network resources and (virtual) network functions isolated and customized over the same physical infrastructure~\cite{Galis:2018}. Such dedicated networks, built on a shared infrastructure can reduce the cost of the network deployment, speeds up the time to market and offer individual networks customizations according to customer requirements so that operators can introduce new market services~\cite{Devlic2017NESMO:Framework}. 

Increased flexibility introduces higher complexity in design and operation of network slices. Keys to avoid the CAPEX and OPEX increase is to automate the full lifecycle phases of a slice: (i) preparation phase, (ii) instantiation, configuration and activation phase, (iii) runtime phase and (iv) decommissioning phase~\cite{3GPP:TR28801:2017}. Besides the automation, other management and orchestration use cases of network slicing are fault management, performance management, and policy management. It is also expected multi-operator coordination management in order to create end-to-end network slices across multiple administrative domains and some level of management to be exposed to the network slice tenant~\cite{Contreras:2018}.

\subsection{Transport Networks}
Optical networks evolved from statically assigned single and multi-mode fiber channels to highly flexible modulation schemes using separate wavelengths. Nowadays, the optical equipment allows prompt wavelength conversion and flexible packet-to-optical setups. Given that agility increase, more programmability is being added to optical networks, for instance through PCE-based architectures for application-based network operations (ABNO)~\cite{RFC7491}. 

Under the flag of Software-Defined Optical Networks~\cite{7503119}, such as those based on OpenFlow extensions, different use cases target transport networks to deliver new approaches on wavelength-based routing and virtualization of optical paths.  Like \gls{pce}, different forms of \gls{sdn} abstractions in optical networks come with a logically centralized entity to program network elements encompassing optical paths. In a wider perspective, logical services are implemented through central controllers as part of a \gls{nso} workflow.
 
Optical transport of traffic across long-range areas, from data centers to end customers as Fiber-to-the-X (e.g., houses FTTH, curbs FTTC, Nodes FTTN), involve different intermediate elements requiring packet-optical conversions and vice-versa. An \gls{nso} envisioned in this scenario of packet-optical integration can take advantage of the knowledge about topology and equipment status, therefore optimizing traffic forwarding according to optical and packet-oriented capabilities. For instance, an \gls{nso} could optimize and aggregate \gls{mpls} \glspl{lsp} inside optical transport networks as part of higher-level service lifecycle goals.

Ongoing work at \gls{mef} aims to standardize \gls{sdwan} \cite{MEF:SDWAN:2017}  as the means to flexibly achieve programmable micro-segmented paths -- based on QoS, security and business policies -- across sites (public or private clouds), using overlay tunnels over varied underlay technologies, such as broadband Internet and MPLS. A service orchestrator is needed to tailor and scale paths on-demand to assure application policies by interfacing a controller that manages programmable edge \gls{sdwan} routers, spanning multiple provider sites. WAN traffic can flow through non-trusting administrative domains in heterogeneous wired/wireless underlay networks with different performance metrics.

\subsection{Cloud Data Centers}

Data Centers have long been upgraded with network virtualization for traffic forwarding and scaling L2 domains, such as VXLAN. Current technologies realize hypervisor tunneling for north--south and east--west traffic in data centers. More importantly, with the advent of operating system-level virtualization (a.k.a containers), even more flexible methods of end-host network virtualization have been deployed in data centers --- there are examples already available in commercial products (e.g., VMWare NSX). In addition, computer virtualization platforms also contain networking extensions/plugins for dynamic networking between servers (e.g., Kubernetes and OpenStack). Those logically programmable network fulfillments derive the properties that concern a \gls{nso}.

The orchestration of cloud resources~\cite{liu2011cloud} has been a longstanding topic of research and actual commercial solutions.  \gls{nso} programmability has been increasingly important to keep isolation in-network and at servers for heterogeneous customers that inhabit public clouds (e.g., Azure, AWS and Google Cloud). For instance, Kubernetes, using kube-proxy, defines networking in Google Cloud via a set of dynamic routes associations between service addresses and bridges' addresses in PODs (servers) hosting containers; ideally, a service is maintained independently of the associated containers host location. Container-based orchestration is a production reality, but many challenges remain open~\cite{7185168}, a number of them related to the seamless integration with network services inside the data centers and across data centers.

Similar concepts of \gls{nso} characteristics already exist to program paths optimizing traffic workloads, high throughput and low latency across data centers and to edge \glspl{cdn} -- best examples being Google B4 and Andromeda \gls{sdn} projects. Therefore, \gls{nso} already plays an essential role in data center networking as it became a pioneer in direct application of \gls{sdn} concepts. 

Lately, research topics in this domain concern integration of multiple cloud environments envisioning different guarantees of \gls{sla} for distinct classes of traffic. As more mobile applications evolve towards accomplishing customers requirements for low latency and high throughput (e.g., virtual and augmented reality), \gls{nso} will play an important role in addressing issues originated from those requirements.  

\subsection{Internet of Things}
According to Gubbi et al.~\cite{Gubbi2013InternetDirections}, \gls{iot} is a network of sensing and actuating devices providing the ability to share information through a unified platform. Such devices or "things" may transmit a significant amount of data over a network without requiring human-to-human or human-to-computer interaction. Its application areas include homes, cities, industry, energy systems, agriculture, and health. Due to the amount of generated-data and its dynamic and transient operational behavior, \gls{iot} will lead to scalability and management issues in the process of transport, processing, and storage of the data in real time~\cite{Mijumbi2016NetworkChallenges}. Besides, the various entities involved need to be orchestrated to convert the data into actionable information~\cite{Consel2017InternetOrchestration}. 

\gls{nso} along with \gls{nfv} and \gls{sdn} allow network services to be automatically deployed and managed. In this scenario, \gls{sdn} is responsible for establishing the network connections, \gls{nfv} provides the management of the network functions, and \gls{nso} govern all deployment process of the end-to-end network service. Such paradigms can help to process and manage significant amounts of IoT-generated data with better network efficiency. The separation between resources and services provided by such technologies enables the isolation and lower impact risks of \gls{iot} on other infrastructures. Also, they can reduce human intervention in the operation of the network, feature that is essential to the achievement of \acrlong{iot}.

The authors in~\cite{Wen2017FogServices} propose an orchestrator for \acrlong{iot} that manages all planes of an \gls{iot} ecosystem. The orchestrator selects resources and deploys the services according to security, reliability, and efficiency requirements. This approach enables an overall view of the whole environment, reducing costs and improving the user experience. Thus, orchestration allows the creation of more flexible and scalable services, reducing the probability of failure correlation between application components. 
\section{Network Service Orchestration}
\label{sec:nso}

\subsection{Definitions}
\label{sec:orchdef}

Various communities differ concerning the meaning, assumptions, and scope of orchestration functions. Thus, it is helpful to begin by reviewing the community understanding to get the main concepts and significance. To this end, we overview the leading organizations and efforts defining the term Orchestration in the context of network service.

\gls{nist} \cite{Bohn2011NISTArchitecture} was one of the first organizations to define the concept of Service Orchestration formally. According to NIST vision, orchestration is a process related to the arrangement, coordination, and management of virtualized infrastructure to provide different cloud services to customers.

A couple of years ago, the term orchestration was adopted by \gls{etsi} in the scope of \gls{nfv}. In \gls{etsi} \gls{nfv}, the meaning of orchestration leads to a vague distinction between orchestration and management. According to~\cite{ETSIISG2018}, orchestration is a set of coordinated processes that automate the management and control of information systems to reach a common goal. However, it emphasizes that orchestration could be provided in multiple functional blocks, no primacy over others. Similarly, the Internet Engineering Task Force (\gls{ietf}) comes up with an orchestration definition closely aligned with \gls{etsi}. 

The \gls{onf} \cite{OpenNetworkingFoundation2016FrameworkNetworks} has formally defined orchestration as usage and selection of resources by orchestrator for satisfying client demands according to the service level. Orchestration is considered as a feature of the \gls{sdn} controller being a key part of SDN architecture. \gls{onf} mentioned that main functions of Orchestration are two-fold. First, orchestration implies to split heavy-loaded service requests into service components. Moreover, it distributes the aforementioned components among supported platforms, creating an integrated end-to-end solution across multiple domains.

The ITU-T Recommendation Y.3300 \cite{InternationalTelecommunicationUnion2014ITU-TNetworking} describes the framework of software defined networking. This recommendation defines that \gls{sdn} functions are programming, orchestrating, controlling and managing network resources. Also, it mentions that orchestration provides automated control and management of network resources. Nevertheless, ITU-T does not clarify the difference between \gls{sdn} functions and orchestration, what causes some confusion.

According to 3GPP Technical Specification 28.801 \cite{3GPP2017TRNetwork}, orchestration is responsible for interpreting and translating a given service request into a configuration of resources (physical and/or virtualized), as needed for service establishment. The configuration of resources may use resource allocation policies or actual available resources. 

In the 5G white paper issued by NGMN \cite{NGMNAlliance2015NGMNPaper}, there is an end-to-end management and orchestration entity which composes the proposed architecture, and it is in charge of translating the service request (business models) into infrastructure resources, beyond managing tasks such as resource scaling and network functions geographic distribution. It is worthwhile noting this proposal is similar to the one presented by \gls{etsi} \gls{nfvo}.  

The \gls{mef}~\cite{MEF:Third:2015} proposes \gls{lso} as a reference architecture for multi-domain orchestration. \gls{lso}, based on network-as-a-service principles, extends the \gls{nfvmano} architecture and creates new capabilities. The orchestration of \gls{lso} refers to "automated service management across multiple operator networks that include fulfillment, control, performance, assurance, usage, security, analytics, and policy capabilities."

In addition to all the above-mentioned leading organizations, there are some works in the literature which also define orchestration. According to \cite{Rostami2016Multi-Domain5G}, orchestration enables programmability for creating and deploying end-to-end network services and dynamic network control through a single interface. Thyagaturu et al. \cite{Thyagaturu2016SoftwareSurvey} address orchestration as the coordination of network services and operations in a higher layer, abstracting the underlying physical infrastructure. The work in \cite{Guerzoni2016Multi-domainApproach} makes a generic definition of orchestration as automated management of complex systems and services.

\subsection{NSO Functionality and Scope}
\label{sec:def}

The purpose of this section is to present the \gls{nso} functionality and scope in an implementation free approach. For that, we review the main functional aspects handled by a \gls{nso}.

\begin{figure}[t!]
  \centering
  \includegraphics[scale=.45]{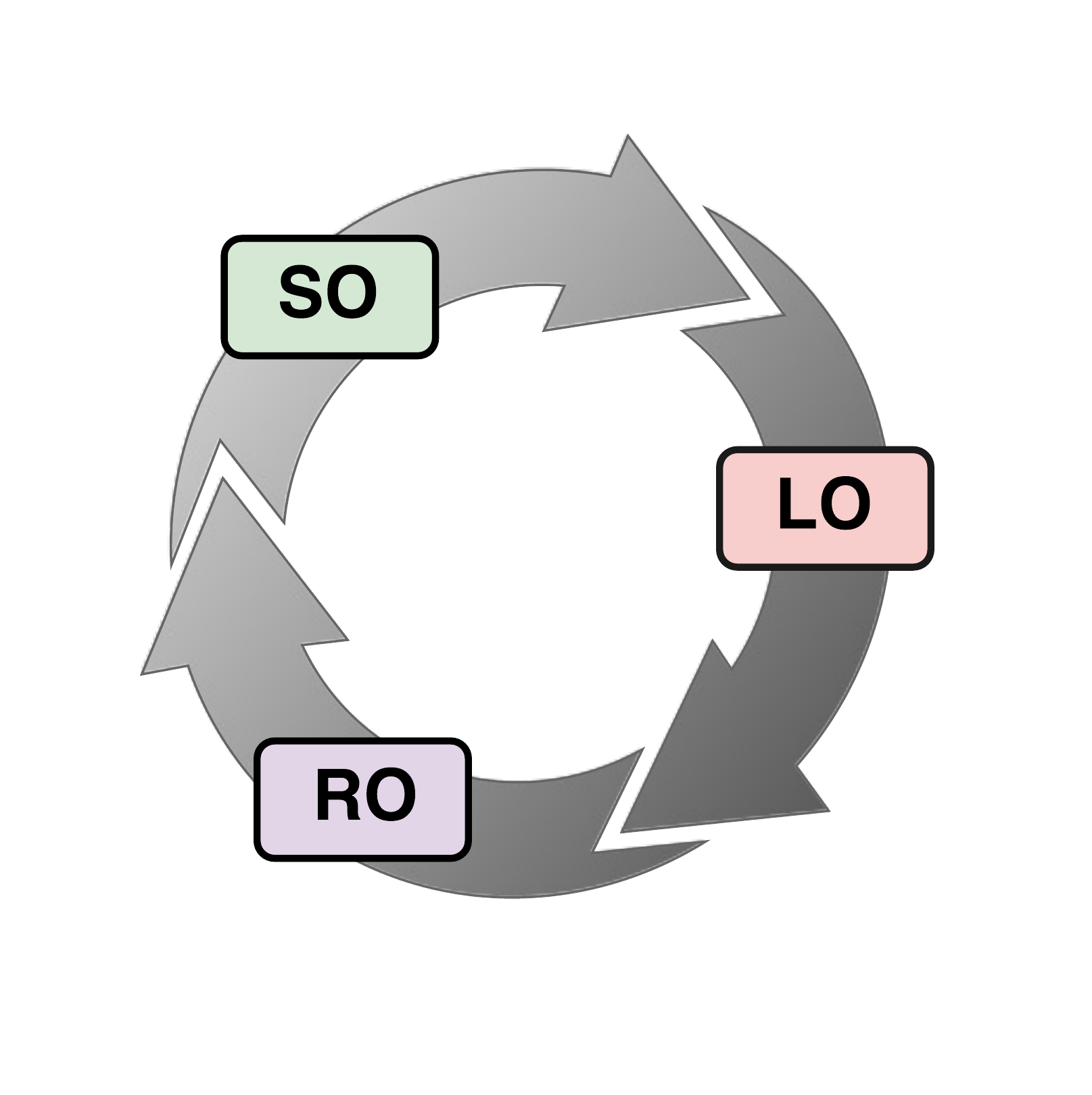}
    \caption{Different orchestrator functions: Resource Orchestration, Service Orchestration, and Lifecycle Orchestration. There is a relationship of dependency and continuity between the functions.}
    \label{funOrch}
\end{figure}

\textbf{Functionalities.} An orchestrator can be classified according to its functional scope: Service Orchestration (SO), Resource Orchestration (RO), and Lifecycle Orchestration (LO). Figure~\ref{funOrch} shows the three primary network service orchestrator functions.

The Service Orchestration is responsible for service composition and decomposition. It can be taken as the upper layer, focused on the interaction with other components such as Marketplace and \gls{oss} / \gls{bss}. The Lifecycle Orchestration deals with the management of workflows, processes, and dependencies across service components. Besides, it maintains the services running according to the contracted Service Level Agreement. Finally, the Resource Orchestration is in charge of mapping service requests to resources, either virtual and/or physical. This mapping occurs across elements such as \gls{nfvo}, \gls{ems}, and \gls{sdn} controllers.

To accomplish this, the orchestrator may be inserted in each layer of telecommunication network stack, from the application layer down to the data plane. Therefore, different orchestrators can exist in each plane, not being limited to a single orchestrator~\cite{Alvizu2016AdvanceEra}. Some of the existing orchestration solutions use an orchestrator logically centralized and consider only ``softwarized'' networks (see Section~\ref{sec:proj}). However, this is very challenging for large and heterogeneous networks. 

\begin{figure*}[thpb]
\centering
 \subfigure[]{
   \includegraphics[scale=.29]{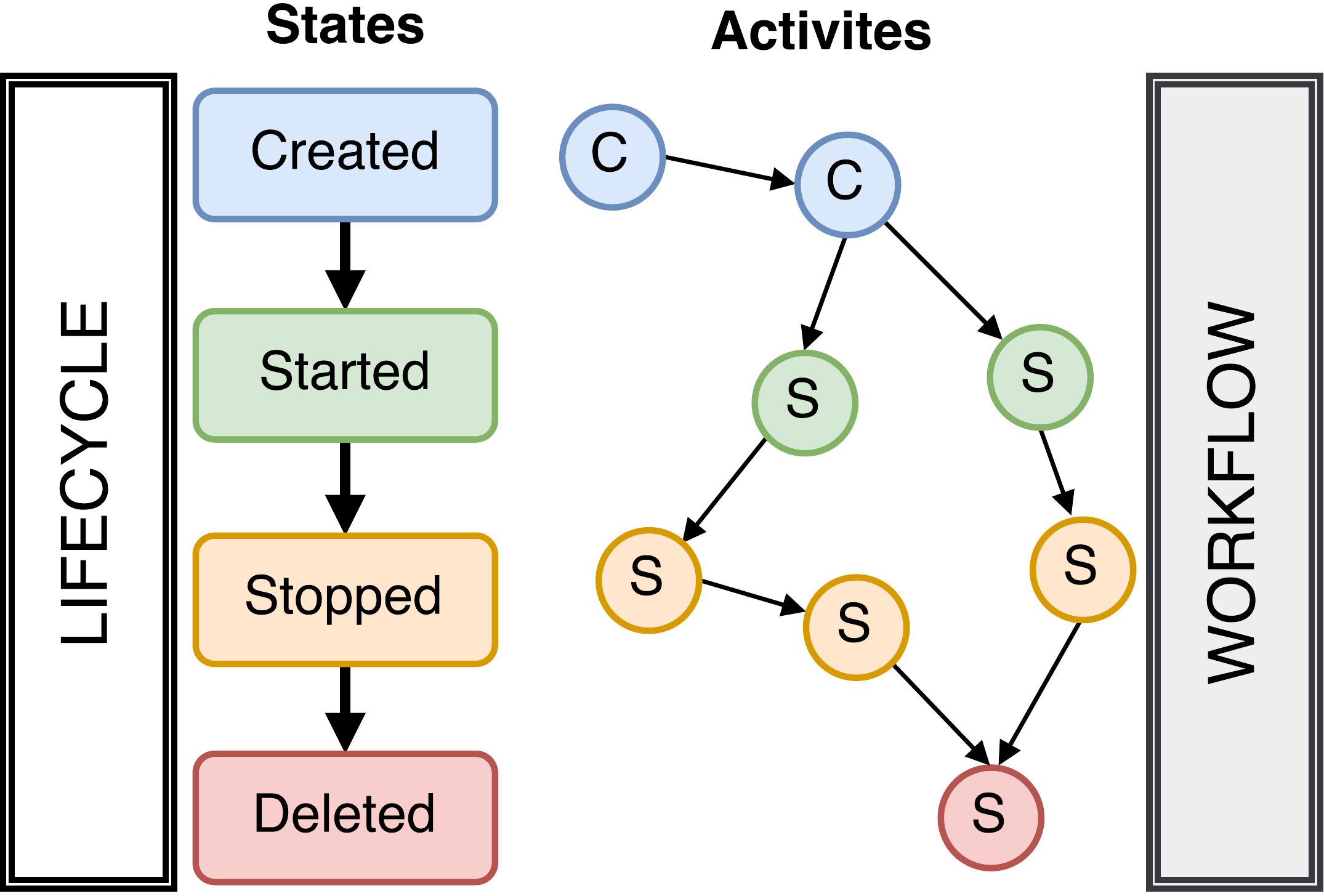}
   \label{fig:lifeWork}
 }
 \subfigure[]{
   \includegraphics[scale=.29]{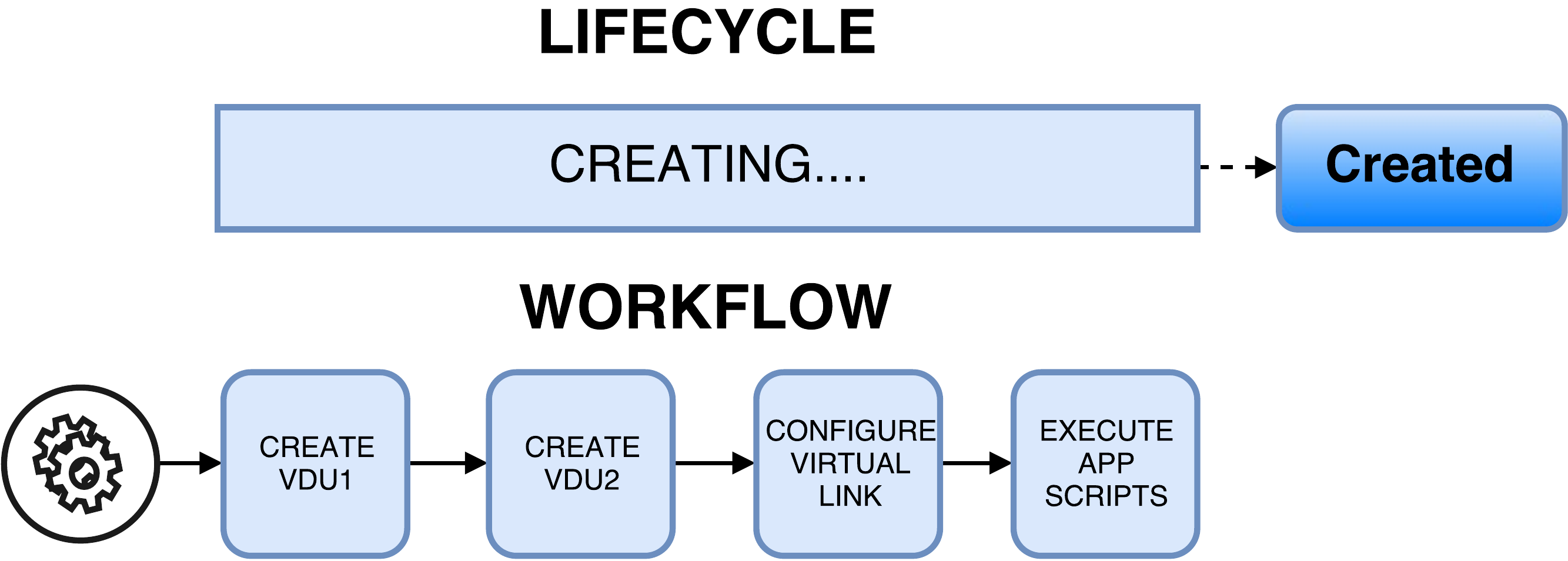} 
   \label{fig:lifeWork2}
 }    
 \caption{Difference between Lifecycle and Workflow: (a) Lifecycle -- sequence of states and workflow -- activities in correct order and (b) example of network service lifecycle.}
 \label{fig:lifeworkflow}  
\end{figure*}

Lifecycle is used to manage a network service with various states (created, provisioned, scaled, stopped, etc.). When some action is applied to a network service (e.g., provision a network service), many activities may be needed to apply to the components of this network service. Hence, a workflow is used to execute a bunch of tasks in the correct order. Each state of lifecycle can generate one or more activities on workflows. The Figure~\ref{fig:lifeWork} depicts the relationship between lifecycle and workflow of a Network Service. 

Figure~\ref{fig:lifeWork2} presents an example to improve the real definition of lifecycle and workflow in the context of network service. One of the states in the service lifecycle is the \textit{Created}. In order to achieve such state is necessary to execute four tasks: create \gls{vdu}1, create \gls{vdu}2, configure network and run the application. Therefore, the state only is changed from creating to created when all those activities are completed.

Service lifecycle automation will allow that requested service remains in a desired state of behavior during its lifetime. With the automation, the system responds proactively to changes network and service conditions without human intervention, getting resilience and faults tolerance. These functional aspects of an orchestrator to guarantee the state of a network towards a service goal are also being referred to as Intent-based Networking (IBN), cf.~\cite{ibn}.

We refer to the \acrfull{nso} when applied in the services deployment performed by telecommunication operators and service providers. We regard NSO not precisely as a unique technology but as a concept to  understand network services in detail, relying on multiple technologies and paradigms to achieve such an overarching goal. In a nutshell, network service orchestration comprises the semantics of requested service, and thereby it coordinates specific actions in order to fulfill the service requirements and to manage its end-to-end lifecycle. 

The entire orchestration process proposed by NSO involves business and operations that go beyond the delivery of \textit{network services} as defined by \gls{etsi}. \gls{etsi} \gls{nfvmano} is a platform for management and orchestration required to provisioning \glspl{vnf} in an \gls{nfv} domain. The \gls{mano} is agnostic and thus has no insight of what is executed within a \gls{vnf}, restricting its responsibility and capability to the VNF instantiation and lifecycle management.

Based on Figure~\ref{mdo}, the \gls{mdo} understands the operating capabilities of the \gls{ns} in a broad sense. When a customer demands an \gls{ns}, firstly it requests the order to a service provider or telecommunication operator through Business-to-Business (B2B) interface or a trading platform we refer to as Marketplace. After that, the \gls{mdo} interacts with any \gls{mano} element or other elements (e.g., OSS/BSS, SDN Controllers, Analytic Systems)  to create the \gls{ns}. Therefore, a given MANO does not know if the VNFs it is deploying is a load balancer, firewall, or gateway. Meanwhile, the \gls{do} just coordinates and manages the orchestration process at a given domain, connecting the involved elements such as network systems, SDN controllers, management software, and IT software platforms.

In this sense, different organizations and telecommunication enterprises have developed many open source projects, driving orchestration evolution towards open standards that will permit the implementation of products with a large scale of integration. Section~\ref{sec:proj} addresses some of these projects.

In addition, the customers are demanding full information regarding a given hired network service such as detailed pricing, real-time analytics, and a precise control over the service. NSO can offer more information to the customers and put more control into their hand. Its objective is to understand the service profoundly and to enable that providers/operators attend customer demands. 

From an operator and service provider viewpoint, NSO enables to set up new end-to-end services in minutes, keeping those services working and ensuring acceptable performance levels. This process reduces OPEX and provides enhanced services creating new market opportunities and raising the revenues.  It opens up chances for different companies to become service providers or provide virtual network functions, as well.

After this analysis, we can identify the main NSO characteristics as follows: 
\begin{itemize}
\item \textit{High-level vision of the \gls{ns}} that permit an overview of all involved domains, technological and administrative. 
\item \textit{Smart services deployment and provisioning}. These are related to in-deep knowledge about the services, what enable better make decisions. 
\item \textit{Single and multi-domain environment support} that provide deployment of end-to-end service independently of geographical location.
\item \textit{Proper interaction with different MANO and non-MANO elements} which leads to better-executed workflows.
\item \textit{Fulfilling new market opportunities}, offering enhanced services and reducing OPEX.    
\end{itemize}

\subsection{Single and Multi-Domain Orchestration}
\label{sec:domain}

The \gls{nso} works at a higher level in the control and management stack with interfaces to the OSS/BSS. During a network service creation, the orchestration process can exceed the domains boundaries being necessary to use resources and/or services of other providers or operators. Such resources comprising physical and virtual components. Thus, the \gls{nso} is supposed to provide service delivery both within single and/or multi-domain environments.

Orchestration in the single and multi-domain environment is different. In a single domain, the orchestrator is in charge of all services and resource availability within its domain as well as has total control over those resources. A domain orchestrator manages the network service lifecycle and interacts with other components not only to control \glspl{vnf}, but also computing, storage, and networking resources. Its scope is limited by administrative boundaries of the provider. As shown in Figure~\ref{mdo}, domain orchestrators can orchestrate heterogeneous technological domains such as \gls{sdn}, \gls{nfv}, Legacy, and Data center. Under a single domain environment, it is noticeable that the domain orchestrator works as described by \gls{etsi} in \cite{ETSIIndustrySpecificationGroupISGNFV2014NetworkOptions}. 

On the other hand, in a multi-domain environment,  local orchestrators do not know the resources and topologies used by other providers. So, multi-domain orchestration is more complex, since it is supposed to provide end-to-end services, which requires cross-domain information exchange features (cf.~\cite{md2}).  
Currently, there is not a standard for information exchange process in multi-domain environments, either multi-technology domains or multiple administrative domains. There are some multi-domain orchestration candidates, e.g., T-NOVA FP7 project \cite{FP7projectT-NOVAT-NOVAInfrastructures}, ONAP~\cite{onap}, Escape \cite{Sonkoly2015Multi-DomainClouds}, and \gls{5gex} \cite{Bernardos20155GInfrastructures}. All of them will be discussed later in this survey.

\begin{figure*}[t!]
\centering
 \subfigure[]{
   \includegraphics[scale=.33]{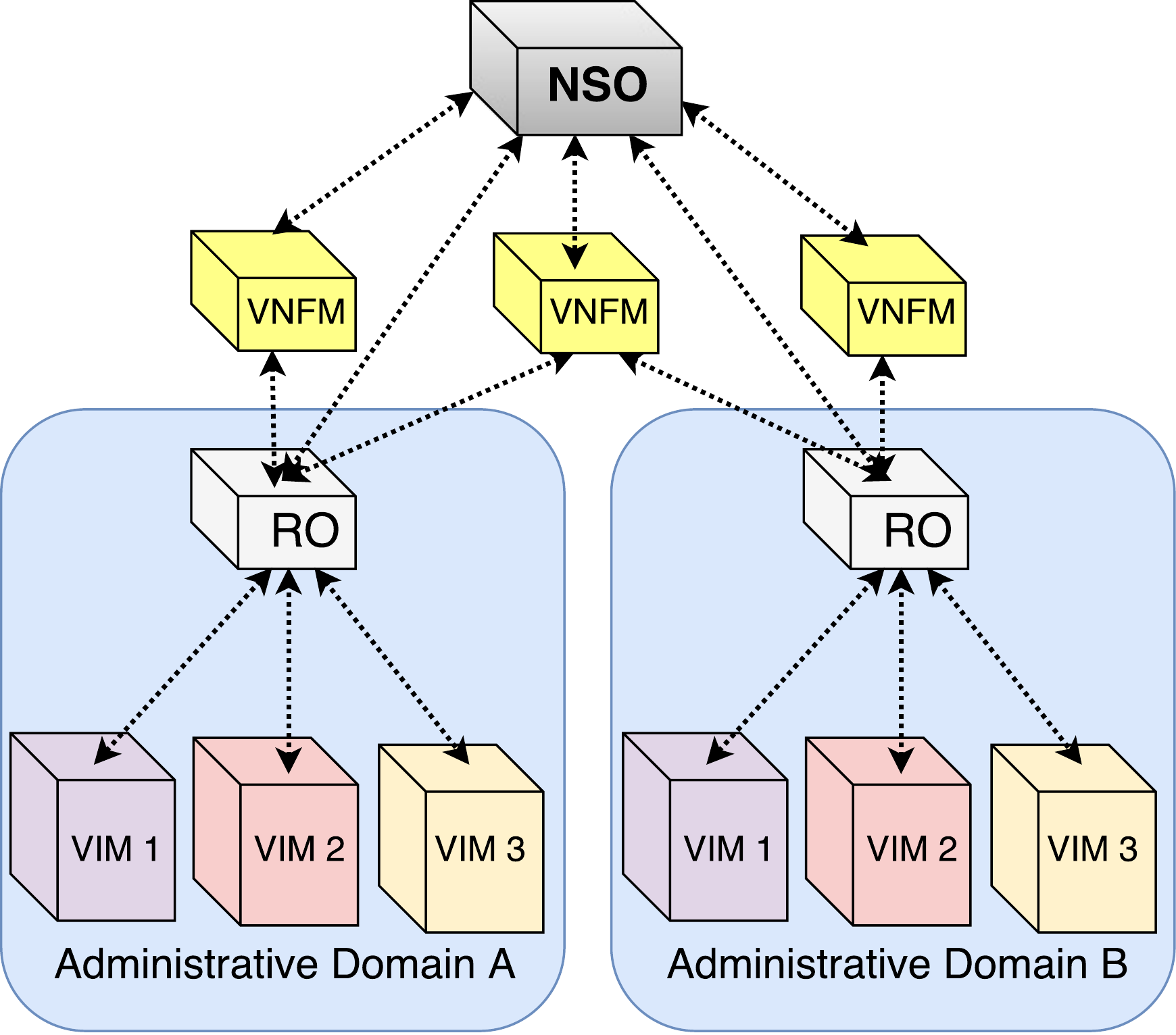}
   \label{fig:use1}
 }
 \subfigure[]{
   \includegraphics[scale=.33]{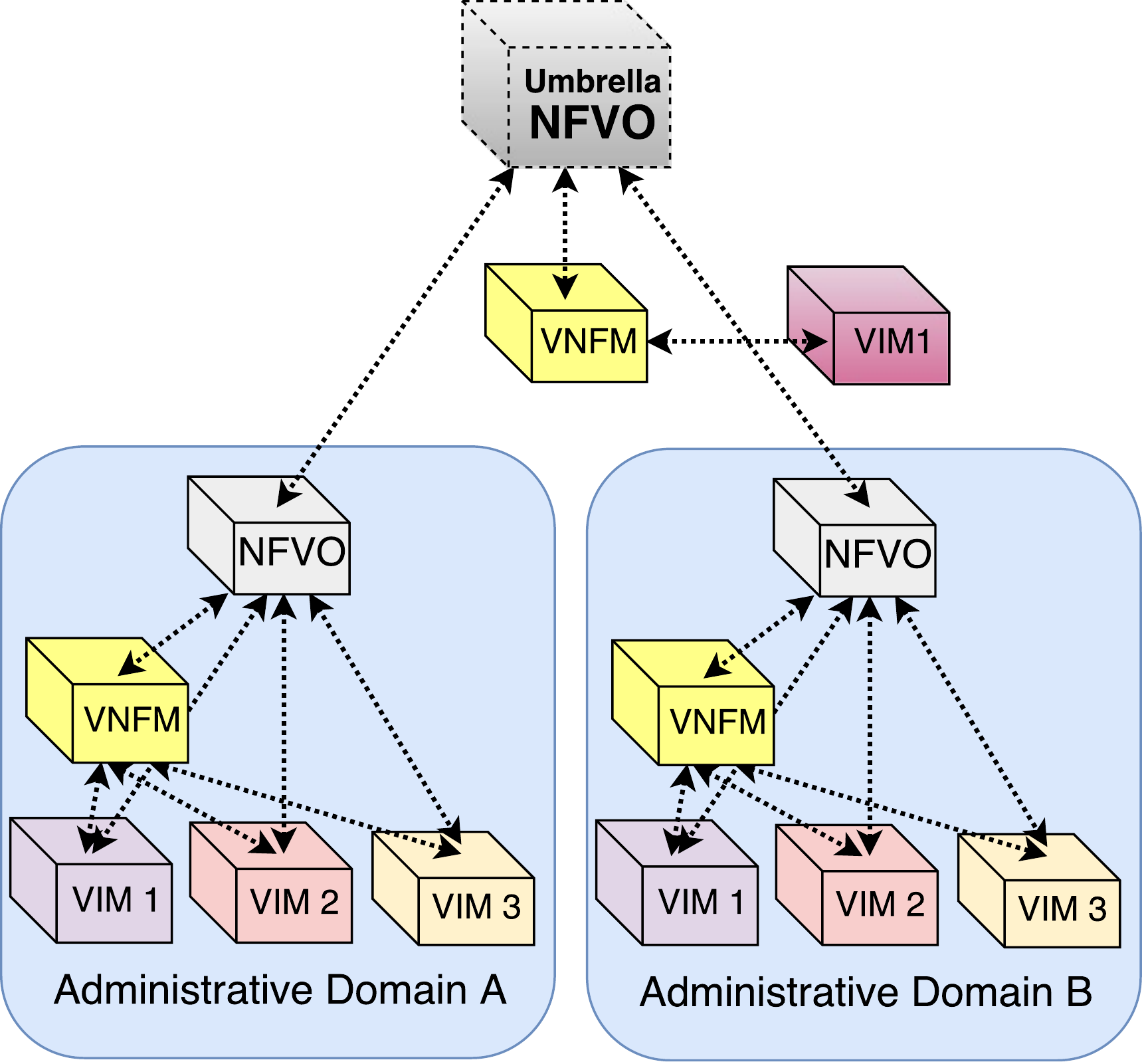} 
   \label{fig:use2}
 }
 \subfigure[]{
   \includegraphics[scale=.33]{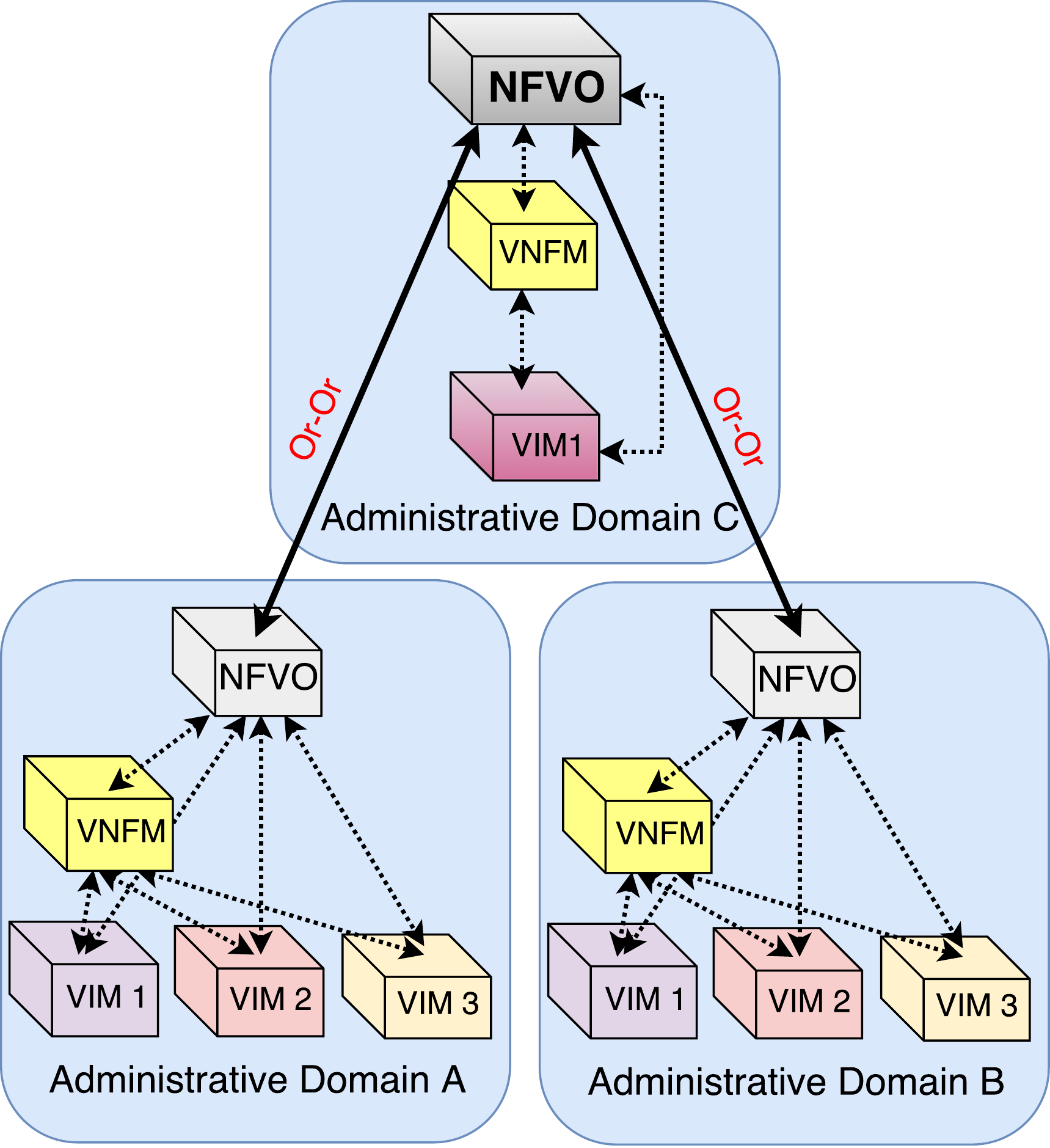} 
   \label{fig:use3}
 }
 \caption{\protect\gls{etsi} approaches for multiple administrative domains: (a) approach in which the orchestrator is split into two components (NSO and RO), (b) approach with multiple orchestrators and a new reference point: Umbrella NFVO, (c) approach that introduces hierarchy and the new reference point Or-Or. Adapted from~\protect\cite{ETSIIndustrySpecificationGroupISGNFV2014NetworkOptions} and~\protect\cite{ETSIGRDomains}.}
 \label{fig:k-clique}  
\end{figure*}

\gls{etsi} proposes some options regarding multi-domain orchestration. Initially, \gls{etsi} \gls{nfv}  Release 2 presents two architectures to address multi-domain scenarios~\cite{ETSIIndustrySpecificationGroupISGNFV2014NetworkOptions}. In the first, the \gls{nfvo} is split into \textit{Network Service Orchestrator}, manages the network service, and \textit{Resource Orchestrator}, provides an abstract resource present in the administrative domain. A use case for this first option is illustrated in Figure~\ref{fig:use1}. A Network Operator offers resources to different departments within the same operator, likewise to a different network operator. One or more Data centers and \glspl{vim} represent an administrative domain and provide an abstracted view of its resources (virtual and physical). The Service Orchestrator and VNF Manager can or cannot be part of another domain. In this use case, service can run on the infrastructure provided and managed by another Service Provider.

The second architecture does not split the \gls{nfvo}, but creates a new reference point between NFVOs (See Figure~\ref{fig:use2}) called  Umbrella \gls{nfvo}. This use case requires the composition of services towards deploying a high-level network service. Such service can include network services hosted and offered by different administrative domains. Each domain is responsible for orchestrating its resources and network services. This approach has objectives similar to first, however, an administrative domain is also composed of \glspl{vnfm} (together with their related \glspl{vnf}) and \gls{nfvo}. The \gls{nfvo} provides standard \gls{nfvo} functionalities, with a scope limited to the network services, \glspl{vnf} and resources that are part of its domain.

More recently, the \gls{etsi} \gls{nfv} Release 3 presented others options to support network services across multiple administrative domains~\cite{ETSIGRDomains}. In particular, the use case entitled ``Network Services provided using multiple administrative domains" proposes a multi-domain architecture using \gls{nfvmano}. Such architecture introduces the new reference point named ``Or-Or" between \glspl{nfvo} to enable communication and interoperability. Differently of the second option (Figure~\ref{fig:use2}), in this approach, there is a hierarchy between the domains. In the example shown in Figure~\ref{fig:use3}, \gls{nfvo} in Administrative Domain C is on-top, using network services offered by Administrative Domains A and B, as well as managing composite \gls{ns} lifecycle.    

In the scope of this paper, end-to-end network services are composed of one or more network functions interconnected by forwarding graphs. Such services might span multiple clouds and geographical locations. Given that, they require complex workflow management, coordination, and synchronization between multiple involved domains (infrastructure entities), which are performed by one (or more) orchestrator(s). Examples of end-to-end services are virtual extensible LAN (VxLAN), video service delivery, and virtual private network.

\subsection{Taxonomy}

\begin{figure*}[thpb]
  \centering
  \includegraphics[scale=.52]{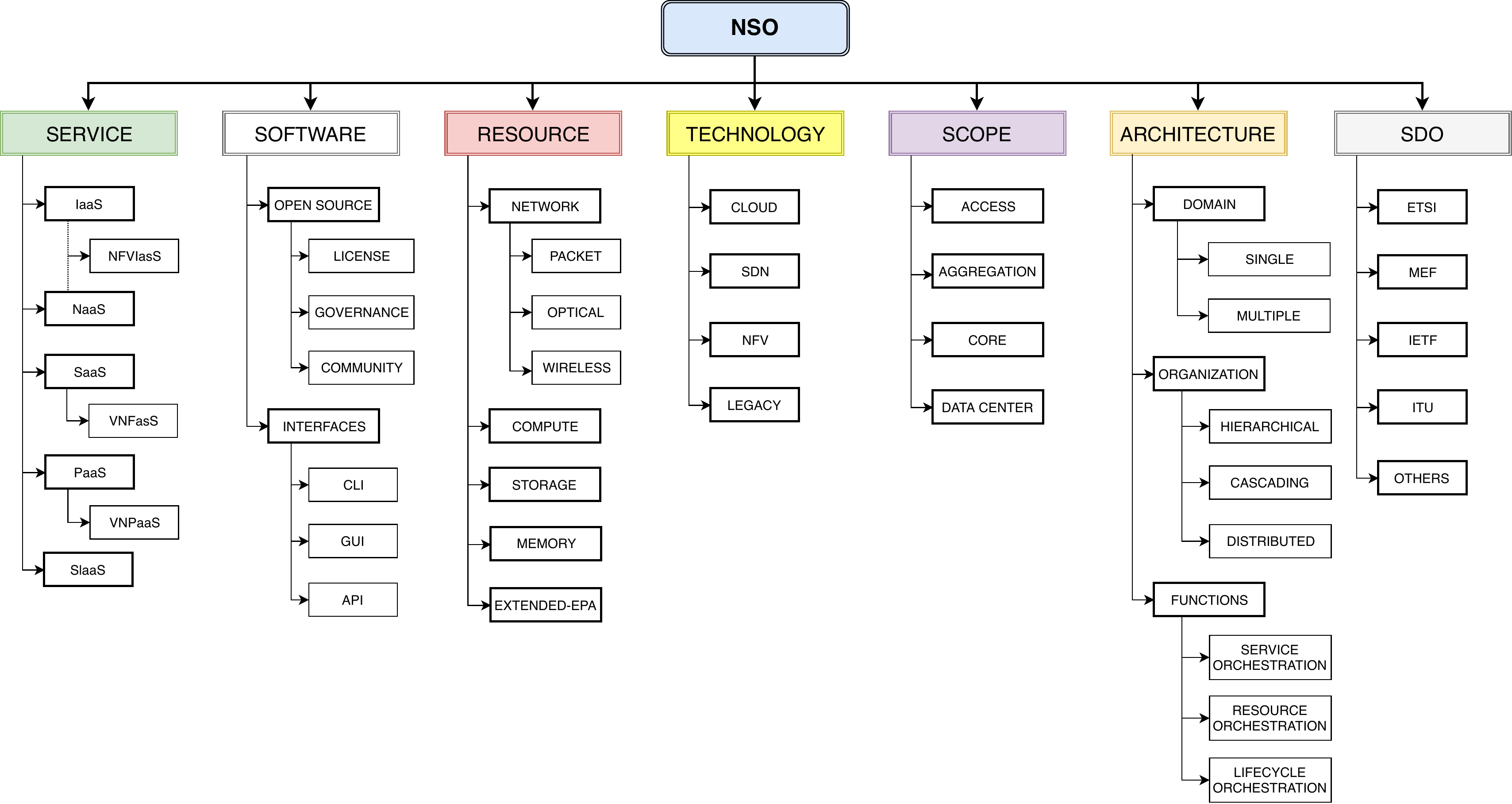}
    \caption{\protect\gls{nso} Taxonomy with seven approach: Service Model, Software, Resource, Technology, Scope of Application, Architecture, and \protect\acrfull{sdo}.}
    \label{tax}
\end{figure*}

While many aspects of orchestration are under active development and commercial roll-outs, others are still in a preliminary maturity phase. This subsection enumerates central concepts and characteristics related to any NSO approach. It becomes very challenging trying to summarize all concepts related to orchestration in a single work, a challenge exacerbated by the fast-evolving pace of so many moving pieces, from standards to enabling technologies. Figure~\ref{tax} presents the proposed taxonomy as the result of extensive literature research as well as practical experiences with a number of orchestration platforms and research projects.   

We identify seven key aspects to characterize network service orchestration: 
\begin{enumerate}
\item \textit{\textbf{Service Models}}. Relates to the type of services unlocked by the \gls{nso}, which may offer new business and relationships and opportunities  (e.g., \gls{vnfaas}, \gls{slaas}).
\item \textbf{\textit{Software}}: Identifies major software-related characteristics of the orchestration solutions, including specificities of the  management and standard interfaces.
\item \textbf{\textit{Resource}}: Refers to the type of underlying resources (e.g., network, compute, and storage) used for the network service deployment.
\item \textbf{\textit{Technology}}: Points to target technologies for \gls{nso} (e.g., Cloud, \gls{sdn}, \gls{nfv}, and Legacy).
\item \textit{\textbf{Scope}}: Considers the application domain in terms of network segments embraced by the network service orchestration (i.e., from access network to data centers).  
\item \textbf{\textit{Architecture}}: Unfolds into three relevant architectural dimensions with relate to single- and multi-domain orchestration and functional  organization.
\item \textit{\textbf{SDO} (Standards Development Organization)}: Relates to standardization activities in scope of the NSO.
\end{enumerate}

Additional sub-areas contribute to an in-depth analysis in different contexts, which are  further  discussed in the following sections. 

\subsubsection{Service Models}
This aspect corresponds to the different service models related to orchestration process. Each service is inserted in the context of cloud, \gls{sdn}, and/or \gls{nfv}. Cloud computing offers three categories of services such as \gls{iaas}, \gls{paas} and \gls{saas} \cite{Leavitt2009}. In \gls{iaas}, Cloud Service Provider (CSP) renders a virtual infrastructure to the customers. In \gls{paas}, CSPs provide development environment as a service. Finally, \gls{saas} is a service that furnishes applications hosted and managed in the cloud. 

\gls{sdn} and \gls{naas} paradigms can be gathered to provide end-to-end service provisioning. While SDN supply the orchestration of underlying network (switches, router, and links), the \gls{naas} is responsible for private access to the network and customer security~\cite{Karakus2017QualitySurvey}. 

The \gls{nfv}, in turn, can offer new services including \gls{nfviaas}, \gls{vnfaas}, \gls{slaas} and \gls{vnpaas}. The \gls{nfviaas} provides jointly \gls{iaas} and \gls{naas} tailored for \gls{nfv}. \gls{vnfaas} is a service that implements virtualized Network Functions to the Enterprises and/or end customers. \gls{vnpaas} is a platform available by service providers allowing customers to create their own network services. The \gls{slaas} is a concept that the slices are traded and used to build infrastructure services.

All these services can work in parallel to offer higher-level services. Each one acts in a specific area and offers features to customers, enterprises, or other providers.   

\subsubsection{Software} 
There are many software artifacts related to orchestration covering from a single cloud environment up to more complex scenarios involving multi-domain orchestration. These solutions are outcomes of open source initiatives, research projects or commercial vendors.   

Open source approaches significantly accelerate consensus, delivering high performance, peer-reviewed code that forms a basis for an ecosystem of solutions. Open source makes it possible to create a single unified orchestration abstraction.  Thus, both research projects and commercial vendors leverage open source technologies to accelerate and improve their solutions. Operators, such as Telefonica, China Mobile, AT\&T, and NTT, appear committed to using open source as a way to speed up their development of orchestration platforms~\cite{Sdxcentral20162016:}.

The \gls{osi}\footnote{http://opensource.org} defines licenses under Open Source Definition compliance, which allows code and software to be freely used, shared and modified. The more popular open source licenses are Apache License 2.0, \gls{bsd}, GNU General Public License (GPL), Mozilla Public License 2.0, and Eclipse Public License. Namely, the most important orchestration projects and frameworks (for instance, Aria, Cloudify, CORD, Gohan, Open Baton, Tacker, ONAP, SONATA, and T-NOVA) present a widespread usage of Apache License 2.0.

Another topic related to open source is governance. In short, governance defines the processes, structures, and organizations. It determines how power is exercised and distributed and how decisions are taken. Commonly, a governing board is responsible for the budget, trademark/legal, marketing, compliance, and overall direction, while a technical steering committee is responsible for technical guidance. 

An open source orchestration project may be organized as a single community (e.g., vendor-lead) or can be hosted (and eventually integrated with other projects) by a foundation entity~\cite{Opensource.comFourOpensource.com}. A remarkable example is the Linux Foundation, which among multiple networking related projects is in charge of  ONAP, an open source platform aiming at the automation, design, orchestration, and management of SDN and NFV services. Another noteworthy example of an orchestration open source project under the Linux Foundation flagship aiming at delivering a standard \gls{nfv}/\gls{sdn} platform for the industry is Open Platform for NFV (OPNFV)~\cite{LinuxFoundation}.

NSO solutions need to perform management tasks such as remote device configuration, monitoring and fault management. Moreover, they require defining an interface of communication between various software components. For this, there are diverse types of management and  standard interfaces such as \gls{cli}, \gls{api}, and \gls{gui}. The \gls{cli} just is used to execute commands directly in the software using remote access via SSH or Telnet. The \gls{api} enables the remote management and interconnection with other softwares through specifics commands. The majority of solutions use REST-based \gls{api}. \gls{gui}, in turn, offers a graphic interface that makes it easier its use.   

\subsubsection{Resource}
During the creation of a network service, the resource orchestration is responsible for orchestrating the underlying infrastructure. Such infrastructure is composed of heterogeneous hardware and software, and different features for hosting and connecting the network services. The resources include compute, storage, network~\cite{Ordonez-Lucena2017NetworkChallenges}, memory, and Extended-\gls{epa}. 

Regarding network, there are three types: packet, optical and wireless (e.g., Wi-Fi, wi-max, and mobile network). Compute, storage, and memory are resources shared among a multitude of network services.  

Resources are shared and abstracted making use of virtualization techniques (e.g., para-virtualization~\cite{4299349}, full virtualization~\cite{4482796}, and containers~\cite{6906035}), defining virtual infrastructures that can be used as physical ones.
For an \gls{nso} solution to be suitable, its virtualized functions must deliver near native (i.e., non-virtualized) performance. For that, EPA capabilities need to be implemented and extended in underlying platform providing highly performant and efficient system. Some examples are (i)\textit{~\gls{numa}}, divide the memory into zones, which are allocated to specific CPUs, (ii) \textit{CPU pinning}, run a particular virtual function’s virtual CPU on a specific physical CPU, (iii)~\textit{\gls{dpdk}}, libraries to accelerate packet processing workloads, and (iv) \textit{Native P4 enabled switches}, provide to programmable pipeline and high-performance forwarding. 

\subsubsection{Technology}
NSO involves complex workflows and different technologies in the orchestration processes touching cloud computing, SDN, NFV, and legacy domains.      

The cloud computing paradigm provides resource virtualization and improves resource availability and usage by means of orchestration and management procedures. This includes automatic instantiation, migration, and snapshot of \glspl{vm}, High-Availability, and dynamic allocation of resources~\cite{ETSI2012NetworkAction}. 

The \gls{sdn} promotes control across network layers and logical centralization of network infrastructure management. Its main functions is to connect the \glspl{vnf} and the \gls{nfvi}-\glspl{pop}. In parallel, the \gls{nfv} technology promotes the network functions programming in order to enable elasticity, automation, and resilience in cloud environments \cite{Rotsos2017NetworkSurvey}. As illustrated in Figure~\ref{nso_rel}, cloud computing, \gls{sdn} and \gls{nfv} are enabler technologies to the \gls{nso}. The NSO must also handle legacy technologies such as MPLS, BGP, SONET / SDH, and WDM. 

\subsubsection{Scope}

Resources of operators under an orchestration application domain can be part of access networks, aggregation networks, core networks, and data centers~\cite{5GPPPArchitectureWorkingGroup2016ViewArchitecture}. The access network is the entry point which connects customers to their service provider. It encompasses various technologies, i.e., fixed access, wireless access (Wi-Fi, LTE, radio, WiMAX), optical, and provide connectivity to heterogeneous services such as mobile network and \gls{iot}. The core network is the central part of a telecommunications network that connects local providers to each other. The aggregation network, in turn, connects the access network to core network. The data center is the local where are localized the computing and storage resources.

The infrastructure is formed by heterogeneous technologies that may be owned by different infrastructure providers. The network service orchestration in this environment is a challenging task. The \gls{nso} must have a view of resources and services since access network until the data center to deploy end-to-end network services. Besides, it is also essential to provide consistent and continuous service, independent of the underlying infrastructure~\cite{5GPPPArchitectureWorkingGroup2016ViewArchitecture}. 

\subsubsection{Architecture}
An NSO architecture can be divided into three sub-categories: (i) \textit{domain}, (ii) \textit{organization}, and (iii) \textit{functions}. The \textit{domain} refers to coverage of the orchestration process in one or more administrative domains: single-domain and multi-domain. In each scenario, orchestration has its peculiarities and challenges.

Single-domain orchestration studies focus on vertical \gls{nfv}/\gls{sdn} orchestration within the same administrative domain. In our definition, an administrative domain can have multiple technological domains, such as \gls{sdn}, \gls{nfv}, and Legacy. The taxonomy is aligned with \gls{etsi} \gls{nfv} architecture that addresses orchestration for \gls{nfv}. The multi-domain orchestration involves the instantiation of network service among two or more administrative domains. It is composed of planes (or layers) with different functions and architecture topology. The multi-domain interfaces are not present in original \gls{etsi} \gls{nfv} architecture

The \textit{organization} refers to the different architectural arrangements of a \gls{nso} solution. We identified three types of organization: hierarchical, cascading and distributed. The hierarchical approach assumes a high-level orchestrator that has visibility of the entire other domains and capable of configuring services across different domains. The service provider facing the customer as a single entry point will maintain relationships with other providers to complete the requested service. According to \cite{Bohn2011NISTArchitecture}, the hierarchical approach is impractical because of scalability and trust constraints.  
Under the cascade model, the provider partially satisfies the service request but complements the service by using resources from another provider. If this provider does not have all the resources, it also can request for another and so on (e.g., a mobile network provider using a satellite provider). In the distributed model, there is not a central actor, and providers request resource and services from each other on a peer-to-peer fashion.

Finally,  \textit{functions}, as discussed in Sec.~\ref{sec:def}, refers to the main tasks developed by network service orchestrator: service orchestration, resource orchestration, and lifecycle orchestration. These functions can be separated or together in the same component of an orchestration framework. This decision depends on how the orchestrator was developed.

\subsubsection{Standards Development Organization (SDO)}
Several Standards Development Organizations, including \gls{etsi}, \gls{mef}, \gls{ietf}, and \gls{itu}, are actively working on a collection of standards in order to define reference architectures, protocols, and interfaces in the scope of the orchestration domain. Besides, other organizations, academic, vendors and industrial are working in parallel with diverse goals. The main efforts within standardization bodies will be outlined next.
\section{NSO and Standardization} 
\label{sec:stand}

\begin{table*}[t]
\scriptsize
\caption{NSO Standardization Outcomes}
\label{Tab:NSO}
\centering
\renewcommand{\arraystretch}{1.3}
\setlength{\arrayrulewidth}{1pt}
\begin{tabular}{c p{3cm} p{3.2cm} m{8.2cm}}
\\
\hline

\textbf{SDO} & \textbf{Working Group} & \textbf{Scope} & \textbf{Outcomes} \\ \hline\hline

 & &  & Service Quality Metrics for NFV Orchestration \cite{ETSIISGNFVGSMetrics} \\
& & & \cellcolor{gray!25} Management and Orchestration Framework~\cite{ETSIIndustrySpecificationGroupISGNFV2013NetworkFramework} \\
& \multirow{-3}{*}{NFV ISG (Initial)} & & Multiparty Administrative domains \cite{ETSIISGNFV2016GRGuidance} \\ \hhline{~-~-}
& & & \cellcolor{gray!25} VNF Architecture and SDN in NFV Architecture~\cite{ETSIISGNFV2014GSArchitecture} \\
& & & Orchestration of virtualized resources~\cite{ETSIISGNFV2017GSSpecification} \\
& & & \cellcolor{gray!25} Functional requirements for Orchestrator~\cite{ETSIISGNFV2017GSSpecification} \\
& & & Lifecycle management of Network Services~\cite{ETSIISGNFV2017GSSpecification} \\
& \multirow{-5}{*}{NFV ISG (Release 2)} & & \cellcolor{gray!25} Network Service Templates Specification \cite{ETSIISGNFV2017GSSpecificationd} \\
\hhline{~-~-}
& & & Policy management~\cite{ETSIISGNFV2017GR3}\\
& NFV ISG (Release 3) & & \cellcolor{gray!25} Report on architecture options to support multiple administrative domains \cite{ETSIGRDomains} \\
\multirow{-10}{*}{ETSI} &  & \multirow{-9}{*}{NFV} &  End-to-end multi-site services management~\cite{ETSIISGNFV2018} \\ \hline

\multirow{2}{*} {MEF} & \multirow{2}{*}{The Third Network} & \multirow{2}{*}{NFV, LSO} & \cellcolor{gray!25} Lifecycle Service Orchestration Vision \cite{MEF:Third:2015} \\ 
& & & LSO Reference Architecture and Framework~\cite{MEF:LSO:2016} \\ \hline

TM Forum & Project & SDN, NFV & \cellcolor{gray!25} ZOOM (Zero-touch Orchestration, Operations and Management)~\cite{TMForumZOOMProject}\\ \hline

& ABNO & SDN & Orchestrate network resources and services~\cite{RFC7491} \\
IETF & SFC & SFC, NFV & \cellcolor{gray!25} SFC Architecture~\cite{Halpern2015} \\ \hline

& & & White Paper: Next Generation Networks~\cite{NGMNAlliance20155GPaper} \\
& & & \cellcolor{gray!25} Network and Service Management including Orchestration~\cite{NGMN:5G:2017} \\
\multirow{-2}{*} { NGMN } & \multirow{-2}{*} {Work Programme} & \multirow{-2}{*} {5G} &  End-to-End Architecture Framework~\cite{NGMNAlliance2018} \\ \hline

& & & \cellcolor{gray!25}  Management and orchestration for next generation network~\cite{3GPP2017TRNetwork} \\
3GPP & S5 & 5G & Management and orchestration architecture~\cite{3gppStudy:28800:2017} \\ \hline

& & & \cellcolor{gray!25} TOSCA for NFV Version 1.0~\cite{OASIS2017TOSCA1.0} \\
\multirow{-2}{*} {OASIS} & \multirow{-2}{*} {TOSCA} & \multirow{-2}{\linewidth} {Resource and Service Modeling} & TOSCA in YAML Version 1.2~\cite{OASIS2017TOSCA1.2} \\ \hline

\multirow{3}{*}{ONF} & \multirow{3}{\linewidth}{Architecture and Framework} & \multirow{3}{*}{SDN} & \cellcolor{gray!25} SDN Architecture~\cite{ONF:SDN:2016} \\ 	
&&& Mapping Orchestration Application to SDN~\cite{ONF:CSO:2017} \\
&&& \cellcolor{gray!25}
Definition of Orchestration~\cite{ONF:Orch:2017} \\ \hline

&&& Report on Standards Gap Analysis in 5G Network \cite{ITU-T2015FGAnalysis} \\
&&& \cellcolor{gray!25} Terms and definitions for 5G network~\cite{ITU-T2017RecommendationNetwork} \\
&&& 5G Network management and orchestration requirements~\cite{ITU-T2017RecommendationRequirements} \\
&&& \cellcolor{gray!25} 5G Network management and orchestration framework \cite{ITU-T2017RecommendationFramework} \\
&\multirow{-5}{*} {ITU-T SG 13} & \multirow{-5}{\linewidth} {5G Network (IMT-2020) and network softwarization} & Standardization and open source activities related to network softwarization~\cite{ITU-T2017ITU-TIMT-2020}\\
\hhline{~-~-}
\multirow{-6}{*} {ITU} & ITU-R & Mobile, radiodetermination, amateur and related satellite services & \cellcolor{gray!25} Framework and overall objectives of the 5G Network~\cite{ITU-R2015RecommendationBeyond}\\ \hline

\end{tabular}
\end{table*}

Interoperability and standardization constitute essential factors of the success of a network service orchestration solution. An important design goal for any new networking paradigm relates to openness of interfaces, especially in order to overcome interoperability issues~\cite{Rotsos2017NetworkSurvey}. 
Several standardization efforts are delivering collections of norms and recommendations to define  architectural guidelines and/or frameworks in addition to standardized protocol extensions to enable NSO. This section presents the main standardization bodies at the \gls{nso} scope. Table~\ref{Tab:NSO} presents a summary of the main SDOs and organizations related to \gls{nso} standardization, as well as the main outcomes produced to date.

\subsection{\acrfull{etsi}}

\gls{etsi} \gls{isg} \gls{nfv} defines the \gls{mano} architectural framework to enable orchestration of \glspl{vnf} on top of virtualized infrastructures. Since 2012, the group provides pre-standardization studies, specification documents and Proof of Concepts (PoCs) in different areas, including management and orchestration. \gls{nfvo} takes a fundamental role in \gls{nfvmano} functional components, as defined in~\cite{GSNFV-MAN001:2014} realizing:
\textit{(i)} the orchestration of infrastructure resources (including multiple \glspl{vim}), fulfilling the Resource Orchestration functions,
\textit{(ii)} and the management of Network Services, fulfilling the network service orchestration functions.

Logically composing \gls{etsi} \gls{nfvo}, \gls{nso} stipulates general workflows on network services (e.g., scaling, topology/performance management, automation), which consequently reach abstracted functionalities in other \gls{mano} components --- lifecycle management of \glspl{vnf} in coordination with \gls{vnfm} and the consume of \gls{nfvi} resources in accordance with \gls{vim} operational tasks.

Currently, \gls{etsi} matures \gls{nfv} in different areas, such as architecture, testing, evolution and ecosystem. Among ongoing topics approached, network slicing report, multi-administrative domain support~\cite{ETSIIndustrySpecificationGroupISGNFV2014NetworkOptions},~\cite{ETSIGRDomains}, context-aware policies, and multi-site services~\cite{ETSIISGNFV2018} highlight important aspects of evolving the \gls{nfv} architectural framework, including possible new \gls{nso} functionalities. 
In the upcoming years, \gls{etsi} is expected to keep playing a driving role represents a path towards realization of concepts built upon the recommendations/reports, as attested by open source projects such as OPNFV and \gls{osm}.

\subsection{\acrfull{mef}}
\acrfull{mef}'s Third Network~\cite{MEF:Third:2015} approaches \gls{naas} comprising agility, assurance and orchestration as its main characteristics to broach \gls{lso} in their defined Carrier Ethernet 2.0. \gls{lso}, as a primer component, provides network service lifecycle management when approaching series of capabilities (e.g., control, performance, analytics) towards fulfilling network service level specifications. \acrfull{mef}'s \gls{lso} provides re-usable engineering specifications to realize end-to-end automated and orchestrated connectivity services through common information models, open \glspl{api}, well-defined interface profiles, and attaining detailed business process flows. Therefore, in \gls{lso} Service Providers orchestrate connectivity across all internal and external domains from one or more network administrative domains. 

A detailed \gls{lso} reference architecture~\cite{MEF:LSO:2016} presents common functional components and interfaces being exemplified in comparison with \gls{etsi} \gls{nfv} framework and \gls{onf} \gls{sdn} architecture. Internally, a Service Orchestration Functionality provides to \gls{lso} coordinated end-to-end management and control of Layer 2 and Layer 3 Connectivity Services realizing lifecycle management supporting different capabilities.
Besides, \gls{lso} defines \glspl{api} for essential functions such as service ordering, configuration, fulfillment, assurance and billing. A recent example of \gls{mef}'s use case conceptualization presents an understanding of \gls{sdwan} managed services in face of \gls{lso} reference architecture~\cite{MEF:SDWAN:2017}. Note that the \gls{lso} functionalities are similar to our \gls{nso} approach.

\subsection{TeleManagement Forum (TM Forum)}
TeleManagement Forum (TM Forum) is a global association for
digital businesses (e.g., service providers, telecom operators, etc.) which provides industry best practices, standards and proofs-of-concept for the operational management systems, also known as Operations Support Systems (OSSs). 

One of the biggest TM Forum achievements is the definition of a telecom business process (eTOM) and application (TAM) maps, including all activities related to an operator, from the services design to the runtime operation, considering assurance, charging, and billing of the customer, among others. In order to accommodate the \gls{sdn}/\gls{nfv} impacts, the TM Forum has created the Zero-Touch Orchestration, Operations and Management (ZOOM) program, which intends to build more dynamic support systems, fostering service and business agility.

As a related research project, SELFNET is, on one side, actively following and aligning its architecture definition with the TM Forum ZOOM and FMO recommendations. Additionally, SELFNET, through one partner of the consortium that is an active member of TM Forum, is also going to actively contribute to the ZOOM working group with respect to the impact that the NFV/SDN paradigm has on the \gls{oss} information model (CFS --– Customer Facing Service, RFS --– Resource Facing Services, LR --– Logical Resources, PR --– Physical Resources). Besides the ZOOM working group, SELFNET will also contribute to the FMO working group by participating in the next generation \gls{oss} architecture, which includes the autonomic management capabilities to close the autonomic management loop: 1) Supervision – 2) Autonomic – 3) Orchestration/Actuation.

\subsection{\acrfull{ietf}}
Different working and research groups at \gls{ietf} address \gls{nso} from varying angles. Traffic Engineering Architecture and Signaling (TEAS) working group characterizes protocols, methods, interfaces, and mechanisms for centralized (e.g., PCE) and distributed path computation (e.g., MPLS, GMPLS) of traffic engineered paths/tunnels delivering specific network metrics (e.g., throughput, latency).  
\gls{abno}~\cite{RFC7491} proposes modular a modular control architecture, standardized by \gls{ietf} aggregating already standard components, such as \gls{pce} to orchestrate connectivity services.  
\gls{sfc} \gls{wg} defines a distributed architecture to enable network elements compute \gls{nf} forwarding graphs realizing overlay paths.
 
The list of protocols involved in NSO is by far not complete and many new extensions to existing protocols and new ones are expected due to the broadening needs of interoperable network service orchestration solutions.

Conceptually, IETF establishes no direct relationship with orchestration, keeping the core concerns majorly on the development of protocols and not their orchestrated operations. However, the development of protocol-related systems, information models and management interfaces by different working groups are actual enablers of  orchestration of services such as layer 3 VPN\footnote{https://datatracker.ietf.org/wg/l3vpn/about/}.
Even more, as the core network service of the Internet, routing detains the capabilities to be managed by orchestration interfaces through the work being performed by the Interface to the Routing System (i2rs) working group.  
For instance, i2rs facilitates ``information, policies, and operational parameters to be injected into and retrieved (as read or by notification) from the routing system while retaining data consistency and coherency across the routers and routing infrastructure, and among multiple interactions with the routing system''.
As such, RFCs developed by different IETF working groups
sit in the scope of SDN, enablers of programmable networks, and therefore, inherit orchestration capabilities.

\subsection{\acrfull{ngmn}}
\gls{ngmn} in~\cite{NGMN:5G:2017} provides key requirements and high-level architecture principles of Network and Service Management including Orchestration for 5G. Based on a series of user stories (e.g., slice creation, real-time provisioning, 5G end-to-end service management), the document establishes a common set of requirements. Among them self-healing, scalability, testing and automation, analysis, modeling, etc. Regarding orchestration functionalities, the presented user stories introduce components (e.g., \gls{sdn} controllers and \gls{etsi} \gls{nfvo}), which execute actions to perform actors goals. For instance, slice creation would be end-to-end service orchestration interpreting and translate service definitions into a configuration of resources (virtualized or not) needed for service fulfillment.  

As part of the initially envisioned 5G White Paper~\cite{NGMNAlliance20155GPaper}, \gls{ngmn} provided business models and use cases based on added values that 5G would bring for future mobile networks. In general, \gls{sdn} and \gls{nfv} components are listed as enablers for operational sustainability that will drive cost/energy efficiency, flexibility and scalability, operations awareness, among other factors for simplified network deployment, operation, and management. Such technology candidates highlight the importance of orchestration capabilities besides the evolution of radio access technologies towards 5G realization.

In addition, the document~\cite{NGMNAlliance2018} defines the requirements necessary that characterize an End-to-End framework. It considers three possible orchestration architecture: (i) Vertical (Hierarchical), that involves processes that ranges from the business level to lower level resource instantiations, (ii) Federated, when the services are provisioned over multiple operators’ networks or over various domains, and (iii) Hybrid (Federated and Vertical), that include characteristics of both federated and vertical orchestration.

\subsection{\acrfull{3gpp}}
Related to the ongoing specification ``Study on management and orchestration architecture of next generation network and service''~\cite{3gppStudy:28800:2017}, 3GPP analyzes its existing architectural management mechanisms in contrast with next generation networks and services in order to recommend enhancements, for instance, to support network operational features (e.g., real-time, on-demand, automation) as evolution from \gls{lte} management. Among the item sets contained in the scope, the specification addresses: the scenario in which the applications are hosted close to the access network; end-to-end user services; and vertical applications, such as critical communications. 
Another ongoing specification, ``Telecommunication management; Study on management and orchestration of network slicing for next generation network''~\cite{3GPP2017TRNetwork}, presents comprehensive 3GPP views on network slicing associated with automation, sharing, isolation/separation and related aspects of \gls{etsi} \gls{nfv} \gls{mano}. In both documents, use cases and requirements cover single and multi-operator services taking into consideration performance, fault tolerance and configuration aspects.

3GPP establishes a relationship with orchestration most notably through management models for network slicing. Related to slicing, the 3GPP TS28 series of documents defines, among other specifications, a network resource model for 5G networks. 
In a protocol and technology neutral way, such models enable management interfaces for the lifecycle management of 5G networks (e.g., core, access, and radio access technologies). 
Closely related to NFV, in 3GPP the study of management and virtualization aspects of 5G networks takes places involving the characterization of performance management and fault supervision.
For instance, in a first stage, TS 28.545 defines the use cases and requirements for fault supervision of 5G networks and network slicing.
Towards synergic studies of 3GPP systems with NFV, there exists on-going work to elaborate further on the energy efficiency control framework defined in TR 21.866 and identify potential gaps concerning existing management architectures, including self-organizing networks and NFV based architectures.
Therefore, related to all the major benefits of introducing NFV paradigms into 3GPP, the management of 5G networks and network slices addresses orchestration in its essence, concerning mostly fault and performance.

\subsection{\acrfull{oasis}}
\gls{oasis} standardizes \gls{tosca} focused on ``Enhancing the portability and operational management of cloud applications and services across their entire lifecycle''. \gls{tosca} Simple Profile in YAML v1.0 was approved as standard in 2016 in a rapidly growing ecosystem of open source communities, vendors, researchers and cloud service providers. Currently, it is in version 1.2~\cite{OASIS2017TOSCA1.2}. Looking forward, \gls{tosca} Technical Committee develops a Simple Profile for \gls{nfv}~\cite{OASIS2017TOSCA1.0} based on \gls{etsi} \gls{nfv} recommendations. 

Logically, \gls{tosca} allows the expressiveness of service to resource mappings via flexible and compoundable data structures, also providing methods for specifying workflows and, therefore, enable lifecycle management tasks. In both Simple and \gls{nfv} Profiles, \gls{tosca} models service behaviors defining components containing capabilities and requirements, and relationships among them. \gls{tosca} realizes a compliant model of conformance and interoperability for \glspl{nso}, enhancing the portability of network services. 

\subsection{\acrfull{onf}}
At \gls{onf}, the \gls{sdn} architecture defines orchestration as TR-521~\cite{ONF:Orch:2017} states: ``In the sense of feedback control, orchestration is the defining characteristic of an SDN controller. Orchestration is the selection of resources to satisfy service demands in an optimal way, where the available resources, the service demands, and the optimization criteria are all subject to change''.  

Logically, \gls{onf} perceives the \gls{sdn} controller jointly overseeing service and resource-oriented models to orchestrate network services through intents on a client-server basis. From top-to-bottom, a service-oriented perspective relates to invocation and management of a service-oriented \gls{api} to establish one or more service contexts and to fulfill client's requested service attributes. 
Such requirements guide the \gls{sdn} controller in orchestrating and virtualizing underlying resources to build mappings that satisfy the network service abstraction and realization. While in a bottom-up view, a resource-oriented model consists of \gls{sdn} controller exposing underlying resource contexts so clients might query information and request services on top of them. In accordance, resource alterations might imply in reallocation or exception handling of service behavior, which might be contained in policies specified by client's specific attributes in a service request.

Recursively, stacks of \gls{sdn} controllers might coordinate a hierarchy of network service requests into resource allocation according to their visibility and control of underlying technological and administrative network domains (e.g., Cross Stratum Orchestration~\cite{ONF:CSO:2017}). Thus, \gls{sdn} controllers might have North--south and/or East/West relationships with each other. At last, a common ground for orchestration concepts is published by \gls{onf} as ``Orchestration: A More Holistic View''~\cite{ONF:Orch:2017}, elucidating considerations of its capabilities, among them, employing policy to guide decisions and resources feedback, as well its analysis.

\subsection {\acrfull{itu}}
International Telecommunications Union (ITU) is the United Nations specialized agency for information and communication technologies (ICTs). It develops technical standards that ensure networks and technologies seamlessly interconnected. The Study Groups of ITU’s Telecommunication Standardization Sector (ITU-T) develops international standards known as ITU-T Recommendations which act as defining elements in the global infrastructure of ICTs~\cite{ITUITU:World}.

The ITU is working on the definition of the framework and overall objectives of the future 5G systems, named as IMT-2020 (International Mobile Telecommunications for 2020) systems~\cite{ITU-R2015RecommendationBeyond}. The documentation is detailed in Recommendation ITU-R M.2083-0. It describes potential user and application trends, growth in traffic, technological trends and spectrum implications aiming to provide guidelines on the telecommunications for 2020 and beyond.

Besides, the Study Group 13 of ITU-T is developing a report on standards gap analysis~\cite{ITU-T2015FGAnalysis} that describes the high-level view of the network architecture for IMT-2020 including requirements, gap analyses, and design principles of IMT-2020. Its objective is to give directions for developing standards on network architecture in IMT-2020. In this report also includes the study areas:  end-to-end quality of service (QoS) framework, emerging network technologies, mobile fronthaul and backhaul, and network softwarization. The report is based on the related works in ITU-R and other SDOs.
\section{Research Projects}
\label{sec:project}

This section presents an overview of relevant \gls{nso} research projects and positions our taxonomy accordingly as summarized in Table~\ref{table:project}, providing a single vision of their scope and status. 
The following subsections are identified by project name and its duration.

\subsection{T-NOVA (2014/01-2016/12)}

The focus of the FP7 T-NOVA project~\cite{FP7projectT-NOVAT-NOVAInfrastructures} is to design and implement an integrated management architecture for the automated provision, configuration, monitoring and optimization of network connectivity and Network Functions as a Service (NFaaS). Such architecture includes: (i) a micro-service based on NFV orchestration platform--called TeNOR~\cite{7502419}, (ii) an infrastructure visualization and management environment and (iii) an NFV Marketplace where a set of network services and functions can be created and published by service providers and, subsequently, acquired and instantiated on-demand by customers
 or others providers.

In the T-NOVA architecture, TeNOR is the highest-level infrastructure management entity that supports multi-pop/multi-administration domain, transport network (i.e.MPLS, Optical, Carrier    Ethernet, etc.) management between POPs, and data center cloud assets. The TeNOR Orchestrator is split into two elements: (i) \textit{Network Service Orchestrator} that manages the Network Service lifecycle, and (ii) \textit{Virtualized Resource Orchestrator} that orchestrates the underlying computing and network resources~\cite{Kourtis2017T-NOVA:Infrastructures}. 

T-NOVA leverages cloud management architectures for the provision of resources (compute and storage) and extends SDN for efficient management of the network infrastructure~\cite{T-NOVAD2.1:Requirements}. Its architecture is based on concepts from ETSI NFV model and expands it with a marketplace layer and specific add-on features. All softwares produced in the project are available as open source at github\footnote{https://github.com/T-NOVA}. Moreover, during the three years of  the  project, other results have led to papers   published in international refereed journals (7) and conferences (31), demos/exhibitions of the developed systems (8), and workshops/meetings with liaised projects (6)~\cite{T-NOVA-D833}.

\subsection{UNIFY (2013/11-2016/04)}

The FP7 Unify\footnote{http://www.fp7-unify.eu/} project dedicated to approaching multiple technology domains to orchestrate joint network services concerning compute, storage and networking. The primary focus set flexibility as its core concern, especially to bring methods to automate and verify network services.

The Unify architecture contains components in a hierarchical composition enabling recursiveness.  At the bottom, a set of Controller Adapters (CAs) interface technology-specific domains (e.g., optical, radio, data center) to abstract southbound \glspl{api} for a typical model of information to define software programmability over a network, compute and storage elements, such as virtualized container, \gls{sdn} optical controller and OpenStack.   
Overseeing CAs, Resource Orchestrators (ROs) define ways to manage the abstracted components of technology-domains specifically. For instance, while an RO for a \gls{sdn} controller orchestrates network flows (e.g., allocating bandwidth and latency), an RO for a cloud orchestrator would be concerned more over orchestrate network jointly with compute and storage resources (e.g., allocating memory and disk). Moreover, managing one or more ROs, a global orchestrator performs network service orchestration in multiple technological domains, understanding the service decomposition and outsourcing specific tasks to ROs.

Altogether, Unify presents a common model of information to interconnect different technological domains, CAs, ROs and global orchestrator. Such YANG model was named Virtualizer, and logically defined configurations following the NETCONF protocol. 
Different demos showcasing joint orchestration of computing and network resources were presented, using the open source orchestrator ESCAPE,\footnote{https://github.com/hsnlab/escape} for instance, modeling \glspl{vnf} over data centers interconnected via an \gls{sdn} enabled network domain.

Following the \gls{onf} \gls{sdn} architecture, Unify demonstrated methods to apply recursiveness across its functional components in order to decompose network services to technological-specific domains. 

\subsection{5GEx (03/2015-03/2018) } 

The 5GEx project aims agile exchange mechanisms for contracting, invoking and settling for the wholesale consumption of resources and virtual network service across administrative domains. 
Formed by a consortium of vendors, operators, and universities, 5GEx allows end-to-end network and service elements to mix in multi-vendor, heterogeneous technology and resource environments.
In such way, the project targets business relationships among administrative domains, including possible external service providers without physical infrastructure resources.

Architecturally, 5GEx addresses business-to-business (B2B) and business-to-customer (B2C) relationships across multi-administrative domain orchestrator that might interface different technological domains. 
Basically, 5GEx extends \gls{etsi} \gls{nfv} \gls{mano} architecture with new functional components and interfaces.
Among its main components, the project defines modules for: topology abstraction; topology distribution; resource repository; \gls{sla} manager; policy database; resource monitoring; service catalog; and an inter-provider \gls{nfvo}.
5GEx currently utilizes outcome resources mostly from projects Unify and T-NOVA, especially joining their open source components into already prototyped demonstrations. 

\subsection{SONATA (07/2015-12/2017)}

With 15 partners representing the telecommunication operators, service providers, academic institutes (among others), the \gls{sonata} project~\cite{sonata} targets to address two significant technological challenges envisioned for 5G networks: (i) \textit{flexible programmability} and (ii) \textit{deployment optimization} of software networks for complex services/applications. To do so, \acrshort{sonata} provides an integrated development and operational process for supporting network function chaining and orchestration~\cite{karl2016devops}. 

The major components of the SONATA architecture consist of two parts: (i) the SONATA \textit{Software Development Kit (SDK)} that supports functionalities and tools for the development and validation of \glspl{vnf} and \gls{ns} and (ii) the SONATA \textit{Service Platform}, which offers the functionalities to orchestrate and manage network services during their lifecycles with a MANO framework and interact with the underlying virtual infrastructure through Virtual Infrastructure Managers (VIM) and WAN Infrastructure Managers (WIM)~\cite{Draxler2017SONATA:Networksb}.

The project describes the use cases envisioned for the SONATA framework and the requirements extracted from them. These use cases encompass a wide range of network services including \gls{nfviaas}, VNFaaS,  v\gls{cdn}, and personal security. One of the use cases consists of hierarchical service providers simulating one multi-domain scenario. In this scenario, \gls{sonata} does not address the business aspects only the technical approaches are in scope. \gls{sonata} intends to cover aspects in the cloud, SDN and NFV domains~\cite{SONATAProject2015D2.2Design}.

Moreover, the project proposes to interact and manage with not only VNFs also support legacy~\cite{SONATAProject2016D2.3Design}. Besides, it describes technical requirements for integrating network slicing in the SONATA platform.  The \gls{sonata} framework complies with the \gls{etsi} \gls{nfvmano} reference architecture~\cite{SONATAProject2016D2.3Design}. 

The results of the project are shared with the community through a public repository\footnote{https://github.com/sonata-nfv} under Apache v2.0 license. Besides, it collaborated to open source solutions such as OSM and OpenStack, and contributed to SDOs such as ETSI, IETF, and ITU-T. Research project as 5G-Picture\footnote{https://www.5g-picture-project.eu/}, NRG-5\footnote{http://www.nrg5.eu/}, 5Gmedia\footnote{http://www.5gmedia.eu/}, and 5G-Transformer had adopted SONATA platform in their implementations. Currently, SONATA is now being enhanced and extended by 5GTANGO\footnote{https://5gtango.eu/} project.

\subsection{VITAL (02/2015-07/2017)}

The H2020 VITAL project~\cite{vital} addresses the integration of Terrestrial and Satellite networks through the applicability of two key technologies such as SDN and NFV. The main VITAL outcomes are (i) the virtualization and abstraction of satellite network functions and (ii) supporting Multi-domain service/resource orchestration capabilities for a hybrid combination of satellite and terrestrial networks~\cite{vitalD23}. 

The VITAL overall architecture is in line with the principal directions established by ETSI ISG NFV~\cite{ETSIIndustrySpecificationGroupISGNFV2013NetworkFramework}, with additional concepts extended to the satellite communication domains and network service orchestration deployed across different administrative domains. This architecture includes, among other, functional entities (NFVO, VNFM, SO, Federation Layer) for the provision and management of the NS lifecycle. In addition, a physical network infrastructure block with virtualization support includes SDN and non-SDN (legacy) based network elements for flexible and scalable infrastructure management.

In terms of dissemination, main achievements of this project include the publication of journal and conference papers (20+), participation in different scientific events (21), collaborations with other national and international research projects (4), and seminars in the academic domain (2)~\cite{vitalD65}. Also, during this project, entire open source software packages have been released, such as OpenSAND\footnote{https://forge.net4sat.org/opensand/opensand} and X-MANO\footnote{https://github.com/ingdestino/x-mano}. X-MANO~\cite{francescon2017x}, for example, is a cross-domain network service orchestration framework with support for different orchestration architectures such as hierarchical, cascading and peer-to-peer. Moreover, it introduces an information model and a programmable network service in order to enable confidentiality and network service lifecycle programmability, respectively.

\subsection{5G-Transformer (06/2017-11/2019)}
The 5G-Transformer Project~\cite{5g-TransformerProject20175GVerticals} consists of a group of 18 companies including mobile operators, vendors, and universities. The objective of the project is to transform current’s mobile transport network into a Mobile Transport and Computing Platform (MTP) based on \gls{sdn}, \gls{nfv}, orchestration, and analytics, which brings the Network Slicing paradigm into mobile transport networks. The project will support a variety of vertical industries use cases such as automotive, healthcare, and media/entertainment. 

Likewise, 5G-Transformer defines three new components to the proposed architecture: (i) \textit{vertical slicer} as a logical entry point to create network slices, (ii) \textit{Service Orchestrator} for end-to-end service orchestration and computing resources, and (iii) \textit{Mobile Transport and Computing Platform} for integrate fronthaul and backhaul networks. The Service Orchestrator is the main decision point of the system. It interacts with others \glspl{so} to the end-to-end service (de)composition of virtual resources and orchestrates the resources even across multiple administrative domains. Its function is similar to our definition of \gls{nso}. Internally the components of the architecture are organized hierarchically, but the end-to-end orchestration of services across multiple domains occurs in a distributed way.

The project is in its third year with some outcomes. In this time, many scientific articles (37+) have been published in peer-reviewed journals, conferences, and workshops, as well as it has registered two Intellectual Property Rights. The development activities are published as open source in GitHub\footnote{https://github.com/5g-transformer/}. However, only Mobile Transport and Computing Platform Code is available. PoCs are scheduled to start in March 2019 including the Automotive, e-Health, Media Provider, Factory 4.0, and Mobile Virtual Network Operator vertical industries.  The proposed solutions are aligned with standards developed by 3GPP and ETSI~\cite{H20205G-TRANSFORMERProject2018}.

\begin{table*}[ht!]
\centering
\scriptsize
\caption{Summary of research projects related to NSO}
\renewcommand{\arraystretch}{1.2}
\setlength{\arrayrulewidth}{1pt}
\label{table:project}
\begin{tabular}{llcccccc}
\\
\hline
\textbf{Class}                            & \textbf{Feature}     & \textbf{T-Nova} & \textbf{Unify} & \textbf{5GEx} & \textbf{SONATA} & \textbf{VITAL} & \textbf{5G-T}   
\\ \hline \hline
                                                                  & IaaS/NVFIaaS & \Circle & \CIRCLE & \CIRCLE & \CIRCLE & \CIRCLE & \Circle \\      
                                                                  \rowcolor{gray!25} \cellcolor{white}  &  NaaS/NVFIaaS & \Circle & \CIRCLE & \CIRCLE & \CIRCLE & \CIRCLE & \Circle \\ 

& SaaS/VNFaaS & \CIRCLE & \Circle & \CIRCLE & \CIRCLE & \CIRCLE & \Circle \\ 
                                                                 \rowcolor{gray!25} \cellcolor{white} & Paaa/VNPaaS  & \Circle & \Circle & \Circle & \Circle &  \Circle & \Circle \\ 
                                                                 
\multirow{-5}{*}{Service} & SlaaS & \Circle & \Circle & \CIRCLE & \Circle &  \Circle & \CIRCLE \\ \hline

\rowcolor{gray!25} \cellcolor{white} Open Source &  & \CIRCLE & \CIRCLE & \LEFTcircle & \CIRCLE & \CIRCLE & \LEFTcircle \\ \hline
 & Packet & \CIRCLE & \CIRCLE & \CIRCLE & \CIRCLE & \CIRCLE & \CIRCLE\\  

\rowcolor{gray!25} \cellcolor{white} & Optical & \CIRCLE & \Circle & \CIRCLE & \Circle & \CIRCLE & \Circle\\  
 
\multirow{-3}{*}{\begin{tabular}[c]{@{}l@{}}Resource/\\ Network\end{tabular}} & Wireless & \CIRCLE & \Circle & \CIRCLE & \CIRCLE & \CIRCLE & \CIRCLE \\ \hline
                                                                 \rowcolor{gray!25} \cellcolor{white} & Compute &  \CIRCLE & \CIRCLE & \Circle & \Circle & \CIRCLE & \Circle \\ 
\multirow{-2}{*}{Resource} & Storage & \CIRCLE & \CIRCLE & \Circle & \Circle & \CIRCLE & \Circle \\ \hline

\rowcolor{gray!25} \cellcolor{white} & Cloud & \CIRCLE & \CIRCLE & \CIRCLE & \CIRCLE & \CIRCLE & \CIRCLE \\  
                                                                  & SDN & \CIRCLE & \CIRCLE & \CIRCLE & \CIRCLE & \CIRCLE & \CIRCLE \\  

\rowcolor{gray!25} \cellcolor{white} & NFV & \CIRCLE & \CIRCLE & \CIRCLE & \CIRCLE & \CIRCLE & \CIRCLE \\ 

\multirow{-4}{*}{Technology} & Legacy & \CIRCLE & \CIRCLE & \CIRCLE & \LEFTcircle & \Circle & \O \\ \hline

\rowcolor{gray!25} \cellcolor{white} & Access & \Circle & \CIRCLE & \CIRCLE & \CIRCLE & \CIRCLE & \CIRCLE \\ 
                                                                  & Aggregation & \CIRCLE & \CIRCLE & \CIRCLE & \CIRCLE & \CIRCLE & \CIRCLE \\  
                                                                 \rowcolor{gray!25} \cellcolor{white} & Core & \Circle & \CIRCLE & \CIRCLE & \CIRCLE & \CIRCLE & \CIRCLE \\ 

\multirow{-4}{*}{Scope} & Data center & \CIRCLE & \CIRCLE & \CIRCLE & \CIRCLE & \Circle & \CIRCLE \\ \hline
 
\rowcolor{gray!25} \cellcolor{white} & Single & \CIRCLE & \CIRCLE & \Circle & \CIRCLE & \CIRCLE & \CIRCLE \\ \cline{2-8} 

\multirow{-2}{*}{\begin{tabular}[c]{@{}l@{}}Architecture /\\ Domain\end{tabular}} & Multiple & \LEFTcircle & \CIRCLE & \CIRCLE & \LEFTcircle & \CIRCLE & \CIRCLE \\ \hline

\rowcolor{gray!25} \cellcolor{white} & Hierarchical & \CIRCLE & \CIRCLE & \CIRCLE & \CIRCLE & \CIRCLE & \LEFTcircle \\ 

 & Cascade & \Circle & \Circle & \Circle & \Circle & \CIRCLE & \CIRCLE \\ 

\rowcolor{gray!25} \cellcolor{white} \multirow{-3}{*}{\begin{tabular}[c]{@{}l@{}}\cellcolor{white}Architecture /\\ \cellcolor{white}Organization\end{tabular}} & Distributed & \Circle & \Circle & \Circle & \Circle & \CIRCLE & \LEFTcircle \\ \hline
                                                                  & \begin{tabular}[c]{@{}l@{}}Service\\ Orchestration\end{tabular} & \CIRCLE & \Circle & \CIRCLE & \CIRCLE & \CIRCLE & \CIRCLE \\ 
                                                                 \rowcolor{gray!25} \cellcolor{white} & \begin{tabular}[c]{@{}l@{}}Resource\\ Orchestration\end{tabular}  & \CIRCLE & \CIRCLE & \CIRCLE & \CIRCLE &  \LEFTcircle & \CIRCLE \\ 

\multirow{-3}{*}{\begin{tabular}[c]{@{}l@{}}Architecture /\\ Functions\end{tabular}} & \begin{tabular}[c]{@{}l@{}}Lifecycle\\ Orchestration\end{tabular} & \CIRCLE & \Circle & \CIRCLE & \CIRCLE & \CIRCLE & \CIRCLE \\ \hline
                                                                 \rowcolor{gray!25} \cellcolor{white} & ETSI & \CIRCLE & \LEFTcircle & \LEFTcircle & \CIRCLE & \CIRCLE & \LEFTcircle \\ 
                                                                  & MEF & \Circle & \Circle & \Circle & \Circle & \Circle & \Circle \\ 
                                                                 \rowcolor{gray!25} \cellcolor{white} & 3GPP & \Circle & \Circle & \Circle & \Circle & \Circle & \LEFTcircle \\ 
                                                                  & NGMN & \Circle & \Circle & \Circle & \Circle & \Circle & \Circle \\

\rowcolor{gray!25} \cellcolor{white} \multirow{-5}{*}{SDO} & Others & \Circle & \LEFTcircle & \LEFTcircle & \Circle & \Circle & \Circle \\ \hline

\multicolumn{8}{l}{\footnotesize\textit{\Circle \quad  Outside the Scope, \qquad \LEFTcircle \quad Partial Scope, \qquad \CIRCLE \quad Within the Scope, \qquad \O \quad Undefined }}\\
                                                                           \end{tabular}
\end{table*}

\subsection{Other Research Efforts}
Further architectural proposals and research contributions can be found in the recent literature.  

Recent research works have addressed the definition of NFV/SDN architectures. Vilalta et al.~\cite{Vilalta2016SDNServices} present and NFV/SDN architecture for delivery of 5G services across multi technological and administrative domains. The solution is different from the \gls{nfv} reference architecture. It consists of four main functional blocks: Virtualized Functions Orchestrator (VF-O), SDN IT and Network Orchestrator, Cloud/Fog Orchestrator and \gls{sdn} Orchestrator. The VF-O is the main component orchestrating generalized virtualized functions such as \gls{nfv} and IoT. Giotis et al.~\cite{Giotis2015} propose a modular architecture that enables policy-based management of \glsdesc{vnf}s. The proposed architecture can handle the lifecycle of \glspl{vnf} and instantiate applications as service chains. The work also offers an Information Model towards map the \gls{vnf} functions and capabilities.

The work in~\cite{Devlic2017NESMO:Framework} proposes a novel network slicing management and orchestration framework. The proposed framework automates service network design, deployment, configuration, activation, and lifecycle management in a multi-domain environment. It can manage resources of the same type such as \gls{nfv}, \gls{sdn} and \gls{pnf}, belonging to different organizational domains and belonging to the same network domain such as access, core, and transport.

The NECOS project\footnote{http://www.h2020-necos.eu} proposes a cloud network slicing approach, also referred to as Lightweight Slice Defined Cloud (LSDC)~\cite{dantas2018necos}  under a Slice as a Service model where a slice provider orchestrates the required resources to enable end-to-end network services across different segments from federated administrative and/or technical domains. The actual high-level  service orchestration is carried by the slice tenant but leverages technology-independent abstractions for control and monitoring of a different set of resources across multiple domains.

Finally, there still exists a large set of NSO related projects sponsored by the European Union Horizon 2020 research and innovation programme in the 5G Infrastructure Public Private Partnership phases 1, 2, and 3\footnote{https://5g-ppp.eu/}.
In different manners, those projects relate to orchestration detaining common relationship with open source projects such as the Open Source MANO (OSM) initiative, further described.
To quote some of them, for instance: 5GCity\footnote{https://www.5gcity.eu/} aims an infrastructure business model for the 5G city networks; 5G-MEDIA\footnote{http://www.5gmedia.eu/} works to integrate media-industry applications with the underlying 5G programmable service platform based on SDN/NFV technologies; and Sat5G\footnote{http://sat5g-project.eu/} intends to establish a plug-and-play satellite infrastructure integrated with 5G connectivity aiming unserved and underserved areas.
Such myriad of projects walk towards the consolidation of 5G ideas into live demonstrations of projects, such as the coordination of cross-border corridors for 5G experimental test beds.
In essence, all of them were established on ground concepts of NSO, extensively based on and contributing to standardization bodies (e.g., ETSI, ONF, 3GPP).
\section{Enabling Technologies and Solutions}
\label{sec:proj}
Some of the existing orchestrating solutions are just tied to a specific networking environment, and moreover, some of them can orchestrate an only limited number of services~\cite{Kuklinski2016DesignOrchestrators}. In this section, an overview of main orchestration frameworks is presented, including open source, proposed and commercial solutions. The projects cover different technologies and domains. 
Table~\ref{tab:NSOsolutions} summarizes the main characteristics of each open source project. The information in this table is organized as follows: leader entities, \gls{vnf} definition, resource domains (Cloud, SDN, NFV, Legacy), NFV MANO functional blocks implemented (NFVO, VNFM, VIM or VIM-support), Management Interface (CLI, API, GUI), and scope (single/multi-domain).

\subsection{Open Source Solutions}

Open Source Foundations such as the Apache Foundation and the Linux Foundation are increasingly becoming the hosting entities for large collaborative open source projects in the area of networking.  
The most important projects are \gls{onos}, \gls{cord}, Open Daylight, OPNFV and, recently, ONAP, formed by the merger of \gls{openo} and ECOMP. All the projects are important to create a well-defined platform for service orchestration.

Note that to 5G network, standardization and open source are essential for fast innovation. Vendors, operators, and communities are betting on open source solutions. Even so, existing solutions are still not mature enough, and advanced network service orchestration platforms are missing~\cite{Katsalis2016Multi-DomainDirections}.

In early 2016, the Linux Foundation formed the \gls{openo} Project to develop the first open source software framework and orchestrator for agile operations of \gls{sdn} and \gls{nfv}. \gls{onos} is also developing an orchestration platform for the \gls{cord} project to provide \gls{xaas} exploiting \gls{sdn}, micro-services and disaggregation using open source software and commodity hardware~\cite{Alvizu2016AdvanceEra}.

Many open source initiatives towards network service orchestration are being deployed and this including operators, \gls{vnf} vendors and integrators. However, these are still in the early stages. We describe next some of these initiatives.

\subsubsection{Cloudify}
Cloudify~\cite{GigaSpaces2015} is an orchestration-centric framework for cloud orchestration focusing on optimization \gls{nfv} orchestration and management. It provides a \gls{nfvo} and Generic-\gls{vnfm} in the context of the \gls{etsi} \gls{nfv}, and can interact with different \glspl{vim}, containers, and non-virtualized devices and infrastructures. Cloudify is aligned with the \gls{mano} reference architecture but not fully compliant. 

Besides, Cloudify provides full end-to-end lifecycle of \gls{nfv} orchestration through a simple TOSCA-based blueprint following a model-driven and application-centric approach. It includes \gls{aria} as its core orchestration engine providing advanced management and ongoing automation.

In order to help contribute to open source \gls{nfvmano} adoption, Cloudify engages in and sponsors diverse \gls{nfv} projects and standards organizations, such as TOSCA specification, \gls{aria}, \gls{onap} and the NATO's DCIS Cube architecture~\cite{dcisCube}.

\subsubsection{ESCAPE}
Based on the architecture proposed by EU FP7 UNIFY project~\cite{unify}, ESCAPE (Extensible Service ChAin Prototyping Environment) is an NFV proof of concept framework which supports three main layers of the UNIFY architecture: (i) service layer, (ii) orchestrator layer and, (iii) infrastructure layer~\cite{csoma2014escape}. It can operate as a Multi-domain orchestrator for different technological domains, as well as different administrative domains. ESCAPE also supports remote domain management (recursive orchestration), and it operates on joint resource abstraction models (networks and clouds)~\cite{sonkoly2015multi}.  

In the current implementation of the process flow in ESCAPE, it receives a specific service request on its REST API of the Service Layer. It then sends the requested Service Function Chains to the Orchestration Layer to map the service components to its global resource view. As a final step, the calculated service parts are sent to the corresponding local orchestrators towards instantiating the service.

\subsubsection{Gohan}
NTT's Gohan~\cite{gohan} is a MANO-related initiative for \gls{sdn} and \gls{nfv} orchestration. The Gohan architecture is based on micro-services (just as the TeNOR implementation) within a single unified process in order to keep the system architecture and deployment model simple. A Gohan service definition uses a JSON schema (both definition and configuration of resources). With this schema, Gohan delivers a called schema-driven service deployment, and it includes REST-based API server, database backend, command line interface (CLI), and web user-interface (WebUI). Finally, a couple of applicable use cases for the NTT's Gohan include to use it (i) in the Service Catalog and Orchestration Layer on top of Cloud services and (ii) as a kind of NFV MANO which manages both Cloud VIM and legacy network devices. 

\subsubsection{ONAP}
Under the Linux Foundation banner, \acrfull{onap}~\cite{onap} resulted from the union of two open source \gls{mano} initiatives (OPEN-O~\cite{Foundation} and OpenECOMP~\cite{ATT2016ECOMPPaper}). The \gls{onap} software platform deploys a unified architecture and implementation, with robust capabilities for the design, creation, orchestration, monitoring and lifecycle management of physical and virtual network functions~\cite{onapwiki}. Also, the \gls{onap} functionalities are expected to address automated deployment and management and policies optimization through an intelligent operation of network resource using big data and \gls{ai}~\cite{onapconvergedigest}.

ONAP is currently being supported and pushed by largest network and cloud operators and technology providers around the world~\cite{onapGuide}. Therefore, ONAP can be used to design, develop, and implement dynamic network services across service provider's network and/or within its own cloud.

\subsubsection{Open Baton}
Built by the Fraunhofer Fokus Institute and the Technical University of Berlin, Open Baton~\cite{openbatongit} is an open source reference implementation of the NFVO based on the ETSI NFV MANO specification and the TOSCA Standard. It allows it to be a vendor-independent platform (i.e., interoperable with different vendor solutions) and easily extensible (at every level) for supporting new functionalities and existing platforms.

The current Open Baton release 4 includes many different features and components for building a complete environment fully compliant with the NFV specification. Among the most important are: (i) a \gls{nfvo} (exposing TOSCA APIs) , (ii) a generic \gls{vnfm} and Juju \gls{vnfm}, (iii) a marketplace integrated within the Open Baton dashboard, (iv) an Autoscaling and Fault Management System and (v) a powerful event engine for the dispatching of lifecycle events execution.

Finally, Open Baton is included as a supporting project in the project named Orchestra\footnote{https://wiki.opnfv.org/display/PROJ/Orchestra}. This OPNFV initiative seeks to integrate the Open Baton orchestration functionalities with existing OPNFV projects in order to execute testing scenarios (and provide feedbacks) without requiring any modifications in their projects.

\subsubsection{Open Source MANO (OSM)}
\gls{etsi} Open Source MANO~\cite{ETSIOpenMANO} is an ETSI-hosted project to develop an Open Source \gls{nfvmano} platform aligned with \gls{etsi} \gls{nfv} Information Models and that meets the requirements of production \gls{nfv} networks. The project launched its fourth release~\cite{Israel2017OSMOverviewb} in May 2018 and presented improvements in closed‐loop capabilities and modeling and networking logic. In addition, this release provides cloud native installation and a new northbound interface, aligned with ETSI NFV specification SOL005~\cite{ ETSIIndustrySpecificationGroupISGNFV2018NetworkNFV}. 

The \gls{osm} architecture has a clear split of orchestration function between Resource Orchestrator and Service Orchestrator. It integrates open source software initiatives such as Riftware as Network Service Orchestrator and GUI, OpenMANO as Resource Orchestrator (\gls{nfvo}), and Juju~\footnote{https://www.ubuntu.com/cloud/juju} Server as Configuration Manager (G-VNFM). The resource orchestrator supports both cloud and SDN environments. The service orchestrator provides \gls{vnf} and NS lifecycle management and consumes open Information/Data Models, such as YANG. Its architecture covers only a single administrative domain.  

\subsubsection{Tacker}
Tacker~\cite{OpenStackFoundation2016} is an OpenStack project to build a generic \gls{vnfm} and a \gls{nfvo} to deploy network services and \glspl{vnf} on a Cloud/\gls{nfv} infrastructure platform (e.g., OpenStack). Tacker is based on \gls{etsi} \gls{mano} architectural framework, which provides a functional stack to orchestrate end-to-end network services using \glspl{vnf}.

The \gls{nfvo} is responsible for the high-level management of \glspl{vnf} and managing resources in the \gls{vim}. The \gls{vnfm} manages components that belongs to the same \gls{vnf} instance controlling the \gls{vnf} lifecycle. The Tacker also does mapping to SFC (Service Function Chain) and supports autoscaling and TOSCA \gls{nfv} Profile (using heat-translator).

The tacker components are directly integrated into OpenStack and thus provides limited interoperability with others \glspl{vim}. It combines the \gls{nfvo} and \gls{vnfm} into a single element nevertheless, internally, the functionalities are divided. Another limitation is that it just works in single domain environments.   

\subsubsection{TeNOR}
Developed by the T-NOVA project~\cite{FP7projectT-NOVAT-NOVAInfrastructures}, the main focus of this Multitenant/Multi NFVI-PoP orchestration platform is to manage the entire \gls{ns} lifecycle service, optimizing the networking and IT resources usage. TeNOR~\cite{7502419} presents an architecture based on a collection of loosely coupled, collaborating services (also know as micro-service architecture) that ensure a modular operation of the system. Micro-services are responsible for managing, providing and monitoring \gls{ns}/\glspl{vnf}, in addition to forcing SLA agreements and determining required infrastructure resources to support an NS instance. 

The proposed architecture is split into two main components: \textit{Network Service Orchestrator}, responsible for NS lifecycle and associated tasks, and \textit{Virtualized Resource Orchestrator}, responsible for the management of the underlying physical resources. To map the best available location in the infrastructure, TeNOR implements service mapping algorithms using \gls{ns} and \gls{vnf} descriptors. Both descriptors follow the TeNOR's data model specifications that are a derived and extended version of the ETSI NSD and VNFD data model.

\subsubsection{X--MANO}
X-MANO~\cite{francescon2017x} is an orchestration framework to coordinate end-to-end network service delivery across different administrative domains. 

X-MANO introduces components and interfaces to address several challenges and requirements for cross-domain network service orchestration such as (i) business aspects and architectural considerations, (ii) confidentiality, and (iii) life-cycle management. In the former case,  X-MANO supports hierarchical, cascading and peer-to-peer architectural solutions by introducing a flexible, deployment-agnostic federation interface between different administrative and technological domains. The confidentiality requirement is addressed by the introduction of a set of abstractions (backed by a consistent information model) so that each domain advertises capabilities, resources, and \glspl{vnf} without exposing details of implementation to external entities. To address the multi-domain life-cycle management requirement, X-MANO introduces the concept of programmable network service based on a domain specific scripting language to allow network service developers to use a flexible programmable Multi-Domain Network Service Descriptor (MDNS), so that network services are deployed and managed in a customized way.

\subsubsection{XOS}
Designed around the idea of Everything-as-a-Service (XaaS), XOS~\cite{peterson2015xos} unifies SDN, NFV, and Cloud services (all running on commodity servers) under a single uniform programming environment. The XOS software structures is organized around three layers: (i) a Data Model (implemented in Django\footnote{https://www.djangoproject.com/}) which records the logically centralized state of the system, (ii) a set of Views (running on top of the Data Model) for customizing access to the XOS services and (iii) a Controller Framework (from-scratch program) is responsible for distributed state management. 

XOS runs on the top of a mix of service controllers such as data center cloud management systems (e.g., OpenStack), SDN-based network controllers (e.g., ONOS), network hypervisors (e.g., OpenVirtex), virtualized access services (e.g., CORD), etc. This collection of services controllers allows the mapping to XOS onto the ETSI NFV Architecture playing the role of a \gls{vnfm}. Using XOS as the \gls{vnfm} facilitates unbundling the gls{nfvo} and enable to control both a set of EMs and the VIM~\cite{xos}.

\begin{table*}[t]
\centering
\rowcolors{2}{gray!25}{}
\renewcommand{\arraystretch}{1.3}
\setlength{\arrayrulewidth}{1pt}
\tiny
\caption{Summary of Open Source NSO Implementations}
\label{tab:NSOsolutions}
\begin{tabular}{p{1.6cm}p{1.5cm}p{1.7cm}|c|c|c|c|c|c|c|c|c|c|c|}
\multirow{2}{*}{Solution} & \multirow{2}{*}{Leader} & \multirow{2}{*}{VNF Definition} & \multicolumn{4}{c|}{Resource Domain}                                                                           & \multicolumn{3}{c|}{MANO}                                                        & \multicolumn{3}{c|}{Interface Management}                                      & \multicolumn{1}{c|}{Multiple }                                 \\
                          &                         &                                 & \multicolumn{1}{l|}{Cloud} & \multicolumn{1}{l|}{SDN} & \multicolumn{1}{l|}{NFV} & \multicolumn{1}{l|}{Legacy} & \multicolumn{1}{l|}{NFVO} & \multicolumn{1}{l|}{VNFM} & \multicolumn{1}{l|}{VIM} & \multicolumn{1}{l|}{CLI} & \multicolumn{1}{l|}{API} & \multicolumn{1}{l|}{GUI} & \multicolumn{1}{l|}{Domains} \\ \hline\hline 
Cloudify~\cite{GigaSpaces2015}                 & GigaSpace               & TOSCA                           &    \ding{51}                         &                          &      \ding{51}                    &                             &     \ding{51}                       &      \ding{51}                      &                          &           \ding{51}                &        \ding{51}                   &         \ding{51}                  &                               \\
ESCAPE~\cite{unify}                    & FP7 UNIFY               & Unify                           &      \ding{51}                       &       \ding{51}                    &       \ding{51}                   &                             &      \ding{51}                      &                           &       \ding{51}                    &        \ding{51}                   &   \ding{51}                        &                          &         \ding{51}                      \\
Gohan~\cite{gohan}                     & NTT Data                & Own                             &      \ding{51}                       &      \ding{51}                    &       \ding{51}                   &     \ding{51}                        &       \ding{51}                    &        \ding{51}                   &                          &       \ding{51}                   &       \ding{51}                   &      \ding{51}                    &                               \\
ONAP~\cite{onap}                      & Linux Foundation        & HOT, TOSCA, YANG                &       \ding{51}                      &     \ding{51}                     &     \ding{51}                     &    \ding{51}                         &      \ding{51}                     &        \ding{51}                   &    \ding{51}                       &      \ding{51}                    &    \ding{51}                      &      \ding{51}                    &        \ding{51}              \\
Open Baton~\cite{openbatongit}                & Fraunhofer / TU Berlin  & TOSCA, Own                      &        \ding{51}                     &                          &        \ding{51}                  &                             &       \ding{51}                    &       \ding{51}                    &                          &      \ding{51}                    &         \ding{51}                 &         \ding{51}                 &                               \\
OSM~\cite{ETSIOpenMANO}                       & ETSI                    & YANG                            &        \ding{51}                     &       \ding{51}                   &      \ding{51}                    &                             &     \ding{51}                      &          \ding{51}                 &          \ding{51}                &        \ding{51}                  &      \ding{51}                    &           \ding{51}               &                               \\
Tacker~\cite{OpenStackFoundation2016}                    & OpenStack Foundation    & HOT, TOSCA                      &        \ding{51}                     &                          &        \ding{51}                  &                             &     \ding{51}                      &       \ding{51}                    &                          &     \ding{51}                     &          \ding{51}                &        \ding{51}                  &                               \\
TeNOR~\cite{7502419}                     & FP7 T-NOVA              & ETSI                            &         \ding{51}                    &        \ding{51}                  &       \ding{51}                   &                             &        \ding{51}                   &                           &                          &                          &        \ding{51}                  &         \ding{51}                 &                               \\
X-MANO~\cite{francescon2017x}                    & H2020 VITAL             & TOSCA                           &                            &                          &         \ding{51}                 &                             &    \ding{51}                       &                           &                          &                          &       \ding{51}                   &   \ding{51}                      &          \ding{51}                     \\
XOS~\cite{peterson2015xos}                       & ON.Lab                  & \multicolumn{1}{c|}{-}          &         \ding{51}                    &      \ding{51}                    &       \ding{51}                   &                             &                           &        \ding{51}                   &                          &                          &    \ding{51}                      &       \ding{51}                   &     \ding{51}      \\ \hline                  
\end{tabular}
\end{table*}

\subsection{Commercial Solutions}

The commercial orchestration solutions market is rising and will be formed by diverse types of companies including new startups, service provider IT vendors, VNF vendors, and the traditional network equipment vendors~\cite{Sdxcentral2016LifecycleOverview}.    

Some software and hardware vendors already offer network orchestration solutions. Below are presented the major commercial products that we consider as mature and robust solutions. All information about the products was got through the vendor's site and technical reports.

Cisco offers a product named Network Services Orchestrator enabled by Tail-f~\cite{CiscoIncNetworkCisco}. It is an orchestration platform that provides lifecycle service automation for hybrid networks (i.e., multi-vendors). Cisco NSO enables to design and deliver services faster and proposes an end-to-end orchestration across multiple domains. The platform deploys some management and orchestration functions such as \gls{nso}, Telco cloud orchestration, \gls{nfvo}, and \gls{vnfm}.    

The Blue Planet SDN/NFV Orchestration platform~\cite{BluePlanet2017BLUESUITE} is a Ciena's solution that provides an integration of orchestration, management and analytics capabilities. It aims to automate and virtualize network service across physical and virtual domains. The platform supports multiple use cases, including SD-WAN service orchestration, NFV-based service automation, and \gls{cord} orchestration.

Another commercial solution is the HPE Service Director of the Hewlett Packard Enterprise. The product is a service orchestration \gls{oss} solution that manages end-to-end service and provides analytics-based planning and closed-loop automation using declarations-based service model. It supports multi-vendor VNF, multi-VIM, various OpenStack flavors, and multiple SDN controllers.

The Oracle Communications Network Service Orchestration solution~\cite{OracleCommunicationsOracleSolution} orchestrates, automates, and optimizes VNF and network service lifecycle management by integrating with BSS/OSS, service portals, and orchestrators. It has two environments to deploy the network services: one design-time environment to design, define and program the capabilities, and a run-time execution environment to execute the logic programmed and lifecycle management. In essence, it plays the roles of the \gls{nfvo}, Telco cloud orchestration, and end-to-end service.  

Ericsson offers some solutions in the scope of the cloud, \gls{sdn} and orchestration. One of them is the Ericsson Network Manager~\cite{EricssonInc.EricssonManager} that provides a unified multi-layer, multi-domain (\gls{sdn}, \gls{nfv}, radio, transport and core) management systems and plays various roles such as \gls{vnfm}, network slicing, and network analytics. Another product is the Ericsson Orchestrator~\cite{EricssonOrchestrator} that supports Resource Orchestration, VNF Lifecycle Management, and End-to-end Service Orchestration. It uses the ETSI NFV-MANO architecture as reference, playing the role of \gls{nfvo}, \gls{vnfm} and \gls{so}. Lastly, Ericsson Dynamic Orchestration~\cite{DynamicOrchestrator} proposes the creation, provisioning, and assurance of E2E services (intra- and inter-domain) in an automated manner implementing the main features of an \gls{nso} platform (flexible and automated operation and holistic orchestration). 

Many of the products mentioned above are often extensions of proprietary platforms. There are few details publicly available, mostly marketing material. The list of commercial solutions is not exhaustive and will undoubtedly become outdated. However, the overview should serve as a glimpse of the expected rise of commercial NSO solutions in the near future as enabling open source technologies and standards mature.
\section{Challenges and Research Opportunities}
\label{sec:challenge}

\gls{nso} promises to improve efficiency when instantiating (day~1) and operating (day~2) network services, but the path ahead is not without  challenges. 
This section provides a discussion on the main challenges and research opportunities for \gls{nso}, including scalability, security, resource modeling, performance, and interoperability.

\subsection{Interoperability}

Typically, operators infrastructures are organized in several domains that differ in geographical locations, management (e.g., legacy or \gls{sdn}), administrative boundaries, and technologies. One of the challenges for service providers is to create and to manage services across unique and proprietary interfaces, making integration and startup difficult tasks to be achieved, as well as increasing the operational costs.  
 
In this scenario, interoperability is essential to enable the deployment of end-to-end network services. Few end-to-end services will be confined within the boundaries of a single domain. They normally encompass a multi-domain orchestration environment composed of providers and vendors with different incentives and business models~\cite{Katsalis2016Multi-DomainDirections}. There is no consensus about how would be the exchanging process between the multiple actors in deployment end-to-end network services. In fact, \gls{etsi} \gls{mano} architecture does not bring any provisioning for this kind of exchange~\cite{ETSIIndustrySpecificationGroupISGNFV2014NetworkNFV}. 

A number of orchestration solutions based on the ETSI MANO architecture have emerged with the objective of proposing a complete orchestration framework. Table~\ref{tab:NSOsolutions} shows notable solutions. Although the progress made by ETSI in defining architecture and interfaces, each solution uses a particular implementation and data model, which makes interoperability difficult to achieve (cf.~\cite{NOn}). As a result, chaining network functions leveraging different solutions for a single network service deployment and operation is currently a very costly proposition regarding development efforts and time-to-market.   

Standardization is a path to enable interoperability of network services between operators and address limitations that arise in the deployment of services, as explained in Section~\ref{sec:stand}. Another parallel track towards interoperability is a broad adoption of software components and broad agreements on APIs along data and information models fueled by re-usable open source artifacts. 

\subsection{Resource and Service Modeling}

Network services need to be efficiently modeled towards deploying resource requirements, configuration parameters, management policies, and performance metrics. Service modeling will enable abstraction of resources and capabilities of underlying layers. It simplifies the understanding of functions and provides a generic way to represent resource and service. 

However, it is a major challenge to translate higher-level policies, which are generated from the resource allocation and optimization mechanisms, into a lower level configuration. Templates and standards should be developed to guarantee automated and consistent translation~\cite{YongLi2015Software-DefinedSurvey}. Besides, the standardization can enable the interoperability and integration of network services templates and addresses limitations arising in the deployment of services in heterogeneous landscape.

There are templates and data modeling languages for \acrfull{nfv} and \acrfull{ns} such as TOSCA, YANG, and HOT. In addition, some organizations propose their approaches to the definition of Network Services, e.g., Open Baton and Gohan.

\gls{etsi} \gls{nfv} \gls{mano} proposes VNF and Network Service descriptors as templates for the definition of functions and services. According to \gls{etsi}, \gls{ns} is defined as a set of \glspl{vnf} and/or \glspl{pnf} interconnected by \glspl{vl} and one or more \acrlong{vnffg}. 

On the other hand, \gls{etsi} \gls{ns} specifies lowest level resources such as CPU, memory, and network, but it does not extend the resource modeling and does not define a data model to the descriptors~\cite{Mijumbi2016ManagementVirtualizationb}. Thus, its approach is driven to single domain environment~\cite{Garay2016ServiceForward}. 

On the other hand, the \gls{ietf} \gls{sfc} provides the ability to define an ordered list of network services, or service functions (e.g., firewalls, load balancers, DPI) connecting them in a virtual chain. However, \gls{sfc} does not describe the underlying resource, since its primary focus is service operation, apart from the forwarding topology. As opposed to ETSI, SFC scope covers multi-domain connections.   

Resource and service modeling in softwarized networks including multi-domain scenarios need further work. This evolution will enable interoperability of network services and the correct mapping between the high-level configuration and the underlying infrastructure. Currently, the interoperability among the diverse orchestration platforms does not exist.

\subsection{Network Service Lifecycle Management}

Network service lifecycle consists in all process for deployment, execution, and termination of a network service. The Network Service Lifecycle Management is fundamental to ensure the correct operation of the service.

Nevertheless, the network services can have specific lifecycle management requirements. For example, an NS can use specific resources as \gls{sriov}~\cite{5416637} and DPDK or need resources across various domains. This type of requirements becomes harder the service deployment.

One possible solution is service lifecycle automation. It enables lifecycle management without human intervention. Automation can be obtained through heuristic algorithms and machine learning techniques. ONAP is working on new closed control loops (e.g., CLAMP --- Closed Loop Automation Management Platform)\footnote{https://github.com/onap/clamp} towards providing automation, performance optimization and Service Lifecycle Management, eventually leveraging network analytics and machine learning assisted decisions.
Nevertheless, many aspects of run-time (day 2) workflow modeling and implementation remain open, with TOSCA extensions and BPMN/BPML approaches~\cite{DBLP:conf/closer/CalcaterraCMT17} undergoing improvements to meet the needs of NSO-based lifecycle automation.

\subsection{Performance and  Service Assurance}

The changes that orchestration technology brings to the telecommunication infrastructures make them increasingly virtualized and software-based. So, performance is a constant challenge in a highly dynamic environment of virtual functions and services.  

This change reflects on enabling technologies. For instance, the \gls{nfv} should meet performance requirements to support, in a standard server, the packet processing, including high I/O speed, fast transmission, and short delays~\cite{YongLi2015Software-DefinedSurvey}. The \glspl{vnf} must achieve a performance comparable to specialized hardware. According to~\cite{Mijumbi2016NetworkChallenges}, some applications require specific capabilities, but virtualization can degrade their performance. This generates a trade-off between performance and flexibility. However, recent advances in CPU and virtualization technologies are overcoming these challenges include \gls{dpdk}~\cite{LinuxFoundationDPDKKit} -- libraries and drivers for fast packet processing, NetVM~\cite{7036139} -- enabling high bandwidth network functions to operate at near line speed, and ClickOS~\cite{Martins:2014:CAN:2616448.2616491} -- minimalist operating system that supports high throughput, low delay, and isolation. Likewise, the document~\cite{ETSIGSPractises} of the \gls{etsi}  provides a set of recommendations on the minimum requirements that the hardware and virtualized layer should have to achieve high performance.

Another question is performance monitoring coupled with Network Services maintenance. Both require a global view of the resources and a unified control and optimization process with various optimization policies running in it. The monitoring is required to avoid the violation of \glspl{sla} in the Service layer. In order to keep NS performance, it is demanded that the system equally performs in different layers. In multi-domain scenarios, this becomes more complex because it is necessary the exchange of information and resources between different organizations/domains~\cite{md2}. 
VNF benchmarking~\cite{7313620} and NS chain profiling~\cite{7956044} coupled to NSO lifecycles and run-time MANO resource allocation and management decisions are potential techniques towards service guarantees and SLA compliance.  

In addition, a better composition between the traffic forwarding and \gls{nf} placement is required towards optimizing the \gls{ns} deployment. The first steps to provide service performance guarantees are to avoid heavily loaded service nodes and to identify bottleneck links. Algorithms and machine learning techniques can archive better results in this composition.   

Thus, how to achieve high performance is an important problem in the research and development of \gls{nso} solutions. Projects within the 5G Infrastructure Public Private Partnership (5G-PPP)~\cite{elayoubi:hal-01488208} are targeting enhanced performance towards better user experience. 

\subsection{Scalability}

Some studies assume that 5G network might connect 50 billion devices until 2020~\cite{Panwar2016ACommunication},~\cite{Evans2011TheEverything}. This growth is due to the emergence of vertical industries such as Internet of Things, Smart Cities, and Sensor Networks. In this scenario, orchestration process requires the ability to handle the growth of networks and services to support the huge amount of connected nodes.

In addition, the network services can be deployed over different domains managed by third parties, infrastructure covering large geographical space and diverse type of resources such as access, transport, and core networks. This environment demands high scalability of the components involved, including orchestrators, controllers, and managers. 

Most current \gls{nso} use cases are just based on deploying a network service in a controlled scenario. Just a use case is not able to check the scalability of the solution. In a production environment, the orchestrator is responsible for orchestrating millions of customers and services at the same time. Hence, scalability is an important feature for \gls{nso} success.

Some orchestration solutions mainly focus on centralized solutions, which pose scalability issues. The works~\cite{Alvizu2016AdvanceEra} and~\cite{Garay2016ServiceForward} suggest different orchestrators involved in the orchestration process of end-to-end network services, not being limited to a single orchestrator. However, there are several particularities on each layer that could be better explored with specific orchestrators, instead of adopting a global orchestrator approach. In this way, we argue that the whole orchestration process can experience better results if split among different actors (or orchestrators). 

A key challenge is therefore to develop an orchestration process that is massively scalable. This process could involve one or more orchestrators, becoming open and flexible enough to address future applications and enable the integration with external components. Orchestration must avoid the congestions and bottlenecks in the management and orchestration plane to handle the requests for network services.
  
\subsection{Security and Resiliency}

Softwarized networks modify the way how services are deployed replacing the hardware-based network service components with software-based solutions~\cite{Draxler2017SONATA:Networksb}. Through technologies such as \gls{sdn} and \gls{nfv}, such network can provide automation, programmability, and flexibility. Generally, it depends on centralized control, which leads to risks to security and resiliency~\cite{Arfaoui2017SecurityDirections}. Thus, new protection capabilities need to be put in place, including advanced management capabilities such as authentication, access control, and fault management. 

Security and resiliency must be considered both in design and operation stages of network services. Typically, the services are deployed first, prior to any efforts regarding security development. However, security must be a mandatory issue, mainly in a highly connected and virtualized environment. 

Service instantiation involves automated processes that add and delete network elements and functions without human intervention. A critical problem is the addition of a malicious node that can perform attacks, catch valuable information and even the disruption of the entire services.      

An essential requirement for a multi-domain orchestration platform is the capability to hide specific details of each domain. This ensures privacy and confidentiality of the domains, preserving capabilities and resources to an external component~\cite{francescon2017x}.

Resilience in main NSO components such as orchestrators, controllers, and managers is also a critical problem because it can impact directly in overall service operation. Besides, open interfaces that support network programmability and \gls{nso} components communication with other external elements such as \gls{oss} and other orchestrators are an open issue and a hot topic in research ~\cite{Ordonez-Lucena2017NetworkChallenges}, ~\cite{Arfaoui2017SecurityDirections},~\cite{7345422}. In the same direction, the 5G-PPP published a white paper~\cite{elayoubi:hal-01488208} suggesting that the orchestration platform must be secure, reliable and flexible.
\section{Conclusions}
\label{sec:Conclusion}

The traditional telecommunication industry is facing multiple challenges to keep competitive and improve the mode network services are designed, deployed and managed. Architectures and enabling technologies such as Cloud Computing, SDN and NFV, are providing new paths to overcome these challenges in a software-driven approach.  \acrfull{nso} is a strategic element to converge various technology domains and provide a broader and more agile network service footprints. 

In this comprehensive survey on network service orchestration, we highlight its growing importance and try to contribute to an overarching understanding of the common concepts and diverse approaches towards practical embodiments of NSO. We present enabling technologies, clarify the definitions behind the term orchestration, review standardization advances, research projects, commercial solutions, and list a number of open issues and research challenges. 

The application of NSO in some scenarios was also presented, where it is possible to sense its potential and understand the motivation behind so much ongoing work. We also observe a growing trend towards the use of open source components or solutions in orchestration platforms; however, the platforms require to evolve until become suitable for production. An important contribution of this work was the definition of a taxonomy that categorizes the leading characteristics and features related to network service orchestration.

Despite the fast pace of this vibrant topic, we expect this survey to serve as a guideline to researchers and practitioners looking into an overview of network service orchestration fundamentals, a reference to relevant related work and pointers to open research questions.










\section*{Acknowledgment}

This research was supported by the Innovation Center, Ericsson S.A., Brazil, grant UNI.62 and Federal Institute of Piaui --- IFPI. The authors would also like to express their gratitude to review contributions from David Moura,  Lucian Beraldo, Nazrul Islam, and Suneet Singh (in alphabetical order) funded by the EU-Brazil NECOS project under grant agreement no. 777067.
The authors are thankful for any feedback to improve the work. Do not hesitate to contact the authors and/or via \textit{github}: 
https://github.com/intrig-unicamp/publications/tree/master/NSO-Survey.

\bibliographystyle{elsarticle-harv} 
\bibliography{bib/refs}

\end{document}